\documentclass[12pt]{article}
\usepackage{amsmath,amsfonts,latexsym,amssymb,mathtools}
\usepackage{cite}
\usepackage[]{hyperref} 
\input xy
\xyoption{all}

\newcommand{\ie}{\textit{i.e.}}

\newcommand{\beq}{\begin{equation}}
\newcommand{\eeq}{\end{equation}}
\newcommand{\be}{\begin{equation}}
\newcommand{\ee}{\end{equation}}
\newcommand{\bea}{\begin{equation}\begin{aligned}}
\newcommand{\eea}{\end{aligned}\end{equation}}

\begin{document}

\vspace*{0.5in}

\begin{center}

{\large\bf A proposal for (0,2) mirrors of toric varieties}

\vspace{0.2in}

Wei Gu, Eric Sharpe

\vspace*{0.2in}

\begin{tabular}{c}
Physics Department \\
Robeson Hall (0435) \\
Virginia Tech \\
Blacksburg, VA  24061 
\end{tabular}

{\tt weig8@vt.edu}, {\tt ersharpe@vt.edu}

$\,$

\end{center}

In this paper we propose (0,2) mirrors for general
Fano toric varieties with special tangent bundle deformations,
corresponding to subsets of toric deformations.
Our mirrors are of the form of (B/2-twisted) (0,2) Landau-Ginzburg models,
matching Hori-Vafa mirrors on the (2,2) locus.
We compare our predictions
to 
(0,2) mirrors obtained by Chen et al for certain examples of toric varieties, 
and find that they
match.
We also briefly outline conjectures for analogous results for hypersurfaces
in Fano toric varieties.
Our methods utilize results from supersymmetric localization,
which allows us to incidentally gain occasional further insights
into GLSM-based (2,2) mirror constructions.  For example, 
we explicitly verify that closed string 
correlation functions of the original
A-twisted GLSM match those of the mirror B-twisted Landau-Ginzburg model,
as well as (0,2) deformations thereof.

\begin{flushleft}
July 2017
\end{flushleft}

\newpage

\tableofcontents

\newpage

\section{Introduction}

Ordinary mirror symmetry has had a long history in string theory.
This paper concerns a heterotic generalization of ordinary
mirror symmetry, sometimes known as (0,2) mirror symmetry.
Whereas ordinary mirror symmetry relates, in simple cases, pairs of
Calabi-Yau spaces $X_1$, $X_2$, (0,2) mirror symmetry relates pairs $(X_1,
{\cal E}_1)$, $(X_2, {\cal E}_2)$, where ${\cal E}_i \rightarrow X_i$
is a holomorphic vector bundle such that ${\rm ch}_2({\cal E}_i) = 
{\rm ch}_2(TX_i)$.

Ordinary mirror symmetry is now well-understood, but (0,2) mirror symmetry
is still under development, and has been for a number of years
(see {\it e.g.} \cite{bsw,blum-sethi,Adams:2003zy,mp,Melnikov:2012hk,Chen:2016tdd,Chen:2017mxp}).  
Many basics have been worked out:
there is a (0,2) version \cite{blum-sethi} of the Greene-Plesser orbifold
construction \cite{Greene:1990ud}, there has been an attempt \cite{Adams:2003zy}
to duplicate GLSM-based dualities \cite{Morrison:1995yh,Hori:2000kt}, 
and for the case that
${\cal E}$ is a deformation of the tangent bundle of a `reflexively plain'
Calabi-Yau hypersurface, there is a (0,2) analogue \cite{mp} of 
Batyrev's construction \cite{batyrev1,borisov1,bb1}.
Furthermore, there is now a (0,2) analogue of quantum cohomology,
known as quantum sheaf cohomology, which has been developed in {\it e.g.}
\cite{Katz:2004nn,Adams:2005tc,Sharpe:2006qd,Melnikov:2007xi,McOrist:2007kp,Kreuzer:2010ph,McOrist:2010ae,Guffin:2011mx,McOrist:2008ji,Donagi:2011uz,Donagi:2011va,Closset:2015ohf,Donagi:2014kos,Lu:2015yqh,Guo:2015caf,Guo:2016suk}.
The present state-of-the-art is that quantum sheaf cohomology has been
computed for toric varieties and Grassmannians with deformations of the
tangent bundle.  (At present, however, a heterotic analogue of
Gromov-Witten invariants \cite{Candelas:1990rm} is not yet known.)

Ideally, one would like to understand mirrors to basic cases such as the
quintic with a tangent bundle deformation.  At present, not even such 
basic examples are understood.
One strategy to construct such mirrors would be to use abelian duality
\cite{Morrison:1995yh,Hori:2000kt} to construct mirrors to toric ambient
spaces, and then standard tricks to extrapolate to conjectures for mirrors
to compact Calabi-Yau hypersurfaces.

Such a strategy was attempted in \cite{Adams:2003zy}, who discovered
that the methods previously applied in \cite{Hori:2000kt} seem to
crucially require (2,2) supersymmetry -- or at least a (0,2) extension
will require new ideas.  As a result, the (0,2) version of
abelian duality is not presently understood. 
Worse, unlike the case when \cite{Hori:2000kt} was written, until recently
there were no known examples of (0,2) Landau-Ginzburg mirrors to Fano spaces,
not even for simple cases such as ${\mathbb P}^1 \times {\mathbb P}^1$,
which complicates efforts to extend abelian duality to (0,2) cases..

As part of a program of better understanding (0,2) mirror symmetry,
one of the authors has been engaged with various collaborators in a program
of constructing such Landau-Ginzburg mirrors for Fano spaces
\cite{Chen:2016tdd,Chen:2017mxp}, to help cut through the difficulties above.  
In those works, mirrors were
constructed for (0,2) GLSMs for products of projective spaces,
toric del Pezzo surfaces, and Hirzebruch surfaces\footnote{
Most Hirzebruch surfaces are not Fano, but as discussed in \cite{Chen:2017mxp},
one expects them to flow to isolated vacua in the IR, so one expects
to be able to use the same techniques to build a mirror to the GLSM,
which is more properly interpreted as the mirror to a different geometric
phase (the UV phase) of the GLSM.
}, with (Euler-type) tangent bundle deformations.  In each case, mirrors
were constructed in a laborious non-systematic piecemeal fashion
by guessing ansatzes and comparing chiral rings and
correlation functions to determine
coefficients -- no systematic formulas applicable to all cases were
produced.

In this paper we propose formulas for 
(0,2) B/2-twisted mirrors to A/2 models on
toric Fano spaces
(and closely related toric varieties), and present corresponding conjectures for
hypersurfaces, for a special class of Euler-type tangent bundle deformations
corresponding to a subset of
`toric' deformations.  (To be clear, we are proposing a
formula for Landau-Ginzburg mirrors, but
we are not claiming to have 
a worldsheet dualization procedure along the lines of 
\cite{Hori:2000kt}.)

We will check that our systematic
construction successfully duplicates results
(for this special class of deformations) for the examples of
toric Fano surfaces described
in \cite{Chen:2016tdd,Chen:2017mxp}.  
The methods we present here will only
apply to a subset of the deformations considered in 
\cite{Chen:2016tdd,Chen:2017mxp}, but will produce mirrors systematically
and quickly, unlike the methods used in
\cite{Chen:2016tdd,Chen:2017mxp} to arrive at the results presented there. 

Our methods will use ideas and results from supersymmetric localization
\cite{Pestun:2007rz}, first applied to two-dimensional GLSMs in 
\cite{Benini:2012ui,Doroud:2012xw}.

We begin in section~\ref{sect:rev22} by quickly reviewing existing
results on GLSM-based mirror constructions in theories with
(2,2) supersymmetry.  In section~\ref{sect:prop02} we describe our
proposal for (0,2) mirrors to toric Fano varieties.  In 
section~\ref{sect:correlation-fns} we describe formal arguments for why
correlation functions match between the original A/2-twisted GLSM and
the mirror B/2-twisted Landau-Ginzburg model.
In section~\ref{sect:exs} we describe several examples, checking that
the predictions of our proposal match existing results worked
out in \cite{Chen:2016tdd,Chen:2017mxp}.
In section~\ref{sect:hyp} we describe how to formally extend these results
to hypersurfaces, following the same pattern that has been followed for
(2,2) mirror symmetry.

Other recent work on two-dimensional (0,2) theories from different
directions includes {\it e.g.} \cite{Schafer-Nameki:2016cfr,Franco1,gp1,Franco:2016qxh, Franco:2016fxm,Lawrie:2016rqe,
Franco:2017cjj,ahhm,ahhm2,Dedushenko:2017tdw,tatar17}.

\section{Review of (2,2) Fano mirrors}
\label{sect:rev22}

Let us quickly review the mirror ansatz for abelian (2,2) GLSMs for 
Fano toric varieties in \cite{Hori:2000kt}.

\subsection{General aspects}

\subsubsection{Basics}

First, we consider a GLSM with gauge group $U(1)^k$ and $N$ chiral superfields,
with charges encoded in charge matrix $(Q_i^a)$.

Following \cite{Hori:2000kt}, the mirror
is 
a theory with $k$ superfields $\Sigma_a$, as many as $U(1)$s in
the original GLSM, and $N$ twisted chiral fields $Y_i$, 
as many as chiral multiplets in the original GLSM, of periodicity
$2 \pi i$,
with superpotential
\begin{equation}   \label{eq:orig22mirrorw}
W \: = \: \sum_{a=1}^k \Sigma_a \left( \sum_{i=1}^N Q^a_i Y_i - t_a\right) \: + \:
\mu \sum_{i=1}^N \exp(-Y_i),
\end{equation}
where $\mu$ is a scale
factor.

In the expression above, the $\Sigma_a$ act effectively as Lagrange
multipliers,
generating constraints
\begin{equation}  \label{eq:mirror-d-terms}
\sum_{i=1}^N Q^a_i Y_i = t_a
\end{equation}
originating with the D terms of the original theory.
We can solve these constraints formally\footnote{
The expressions given here are entirely formal, and there can be subtleties.
For example, if the entries in $V_i^A$ are fractional, then as is well-known,
the mirror
may have orbifolds. 
} by
writing
\begin{equation}
Y_i \: = \: \sum_{A=1}^{N-k} V_i^A \theta_A \: + \: \tilde{t}_i
\end{equation}
where $\theta_A$ are the surviving physical degrees of freedom,
$\tilde{t}_i$ are solutions of
\begin{equation}  \label{eq:t-ttilde}
\sum_{i=1}^N Q_i^a \tilde{t}_i \: = \: t_a,
\end{equation}
and $V_i^A$ is a rank-$(N-k)$ matrix solving
\begin{equation}
\sum_{i=1}^N Q_i^a V_i^A \: = \: 0.
\end{equation}
(The rank requirement goes hand-in-hand with the statement that
there are $N-k$ independent $\theta_A$'s.)
The periodicity of the $Y_i$'s will lead to interpretations of the
space of $\theta_A$'s in terms of LG orbifolds and character-valued
fields, as we shall review later. 
Note that for $t_i$, $V_i^A$ satisfying the equation above,
\begin{displaymath}
\sum_{i=1}^N Q_i^a Y_i \: = \: \sum_{i} Q_i^a \left( \sum_A V_i^A \theta_A
+ \tilde{t}_i \right) \: = \: t_a,
\end{displaymath}
and so the $V_i^A$ encode a solution of the D-term constraints.

After integrating out the Lagrange multipliers, the superpotential
can be rewritten as
\begin{equation}   \label{eq:22mirrorw-basic}
W \: = \: \mu \sum_{i=1}^N \left( e^{\tilde{t}_i} \prod_{A=1}^{N-k} 
\exp(-V_i^A \theta_A)
\right).
\end{equation}

In this language, the (2,2) mirror map between A- and B-model operators is
(partially) defined by
\begin{equation}  \label{eq:22mirror-basic}
\sum_{a=1}^k Q_i^a \sigma_a  \: \leftrightarrow \: \mu \exp(-Y_i)
\: = \: \mu e^{\tilde{t}_i} \prod_{A=1}^{N-k} \exp(-V_i^A \theta_A),
\end{equation}
which can be derived by differentiating~(\ref{eq:orig22mirrorw}) 
with respect to $Y_i$.
(See for example \cite{Hori:2000kt}[section 3.2], where this is derived as
the equations of motion of the mirror theory.  In the next section,
we will also see that this map is consistent with axial R symmetries.)
In fact, this overdetermines the map -- only a subset of the $Y_i$'s will
be independent variables solving the constraints~(\ref{eq:mirror-d-terms}).  
As we will see
explicitly later, the redundant equations are equivalent to chiral
ring relations (as must follow since they all arise as the same equations of
motion in the mirror), and are also specified by the equations of motion
derived from the superpotential $W$ above.

In appendix~\ref{app:alt} we will briefly outline a variation on the usual
GLSM-based mirror derivation.
Regardless of how the B-model mirror superpotential is obtained,
it can be checked by comparing
closed-string
A model correlation functions between the mirror and the original A-twisted
GLSM using supersymmetric localization.
For (2,2) theories, this can be done at arbitrary genus using the
methods of \cite{Nekrasov:2014xaa,Benini:2015noa},
whereas for (0,2) theories, we can only apply
analogous tests at genus zero.  We will perform such correlation function
checks later in this paper.

\subsubsection{R charges}

Let us take a moment to consider R charges.
In the A-twisted theory, the axial R-charge is in general
broken by nonperturbative
effects, so that under an axial symmetry transformation,
anomalies induce a shift in the theta angle\footnote{
This should not be confused with the fundamental field $\theta_A$
defined earlier.
} by
\begin{displaymath}
\theta^a \: \mapsto \: \theta^a + 2 \alpha \sum_i Q_i^a,
\: \: \:
t_a \: \mapsto \: t_a + 2 i \alpha \sum_i Q_i^a,
\end{displaymath}
for $\alpha$ parametrizing axial R symmetry rotations.
The shift above can formally be described as
\begin{displaymath}
\tilde{t}_i \: \mapsto \: \tilde{t}_i + 2 i \alpha,
\end{displaymath}
(using the relation between $\tilde{t}_i$ and $t^a$ 
in~(\ref{eq:t-ttilde})).
In the same vein,
under the same axial R symmetry, the mirror field $Y_i$ transforms as
\begin{displaymath}
Y_i \: \mapsto \: Y_i + 2 i \alpha,
\end{displaymath}
so that $\exp(-Y_i)$ has axial R-charge $2$.  If we take $\Sigma_a$ to also
have axial R-charge $2$, then it is easy to verify that the entire
mirror superpotential~(\ref{eq:orig22mirrorw}) 
has axial R-charge $2$, as desired,
taking the $t$'s to have nonzero R-charge as described.
In addition, the operator mirror map~(\ref{eq:22mirror-basic}) is also
consistent with axial R-charges in that case.

\subsubsection{Twisted masses}

One can also consider adding twisted masses.  Recall that a twisted
mass can be thought of as the vev of a vector multiplet,
gauging some flavor symmetry.  Taking the vev removes the gauge
field, gauginos, and auxiliary field, and replaces them with a single
mass parameter $\tilde{m}$, corresponding to the vev of the $\sigma$ field.
In the notation of \cite{Witten:1993yc}[equ'n (2.19)],
this means, for a single 
$U(1)$ flavor symmetry that acts on a field $\phi_i$ with
charge $Q_{F,i}$, we add terms to the action of the form
\begin{displaymath}
-2 | \tilde{m}|^2 \sum_i Q_{F,i}^2 | \phi_i |^2 \: - \: \sqrt{2} \sum_i
Q_{F,i} \left( \overline{\tilde{m}} \overline{\psi}_{+, i} \psi_{-, i} +
\tilde{m} \overline{\psi}_{-,i} \psi_{+,i} \right).
\end{displaymath}

In the present case, for a toric variety with no superpotential,
there are at least as many flavor symmetries as chiral superfields modulo
gauged $U(1)$s, {\it i.e.} at least $N-k$ $U(1)$ flavor symmetries.  
(There can also be nonabelian components.)
For simplicity,
we will simply allow for a twisted mass $\tilde{m}_i$ associated
to each chiral superfield, and will not try to distinguish between those
related by gauge $U(1)$s.

Including twisted masses $\tilde{m}_i$,
the full mirror superpotential (before integrating out $\Sigma$'s) takes
the form
\begin{equation}   \label{eq:full22mirrorw}
W \: = \:  \sum_{i=1}^N \left( \sum_{a=1}^k \Sigma_a Q_i^a + \tilde{m}_i
\right) \left( Y_i - \tilde{t}_i \right)
\: + \:
\mu \sum_{i=1}^N \exp(-Y_i).
\end{equation}
This expression manifestly has consistent axial R-charge $2$
(using the `modified' R-charge that acts on $\tilde{t}_i$).
It differs from the more traditional expression 
\cite{Hori:2000kt}[equ'n (3.86)]
\begin{equation}
W \: = \: \sum_{a=1}^k \Sigma_a \left( \sum_i Q_i^a Y_i - t_a \right)
 + \sum_{i=1}^N \tilde{m}_i Y_i
\: + \:
\mu \sum_{i=1}^N \exp(-Y_i),
\end{equation}
by a constant term (proportional to $\sum_{i=1}^N \tilde{m}_i \tilde{t}_i$),
and so defines the same physics.

After including twisted masses, the operator mirror map becomes
\begin{displaymath}
\sum_{a=1}^k Q_i^a \sigma_a \: + \: \tilde{m}_i
\: \leftrightarrow \:  \mu \exp(-Y_i). 
\end{displaymath}
Note that both sides of this expression are consistent with the (modified)
R-charge assignments described above.

Generically in this paper we will absorb $\mu$
into a redefinition of the $Y_i$'s, and so not write it explicitly,
but we mention it here for
completeness. 

Finally, we should remind the reader that in addition to the superpotential
above, one may also need to take an orbifold to define the theory, as is
well-known.  This will happen if, for example, some of the entries in
$(V_i^A)$ are fractions, in order to reflect ambiguities in taking the roots
implicit in resulting expressions such as $\exp(-V_i^A \theta_A)$.

\subsection{Example with twisted masses}

To give another perspective, in this section we will review the (2,2)
mirror to the GLSM for Tot(${\cal O}(-n) \rightarrow
{\mathbb P}^2$), for $n \leq 3$ (and no superpotential), 
and to make this interesting,
we will include twisted masses $\tilde{m}_i$, correspnding to phase
rotations of each field.

The charge matrix for this GLSM is
\begin{displaymath}
Q \: = \: (1,1,1,-n),
\end{displaymath}
and following the usual procedure, the D terms constrain the dual
(twisted) chiral superfields as
\begin{displaymath}
Y_1 + Y_2 + Y_3 - n Y_p \: = \: t.
\end{displaymath}

The standard procedure at this point is to eliminate $Y_p$, and write
the dual potential in terms of $Y_{1-3}$, taking a ${\mathbb Z}_n$ orbifold
to account for the fractional coefficients of the $Y_i$ and its periodicity.  
In other words,
\begin{displaymath}
Y_p \: = \: \frac{1}{n}\left( Y_1 + Y_2 + Y_3 - t \right),
\end{displaymath}
hence the (2,2) superpotential is given by
\begin{eqnarray*}
W & = & \sum_i \tilde{m}_i Y_i +
\exp(-Y_1) + \exp(-Y_2) + \exp(-Y_3) + \exp(-Y_p), \\
& = & \sum_i \tilde{m}_i Y_i +
(\exp(-Y_1/n))^n + (\exp(-Y_2/n))^n + (\exp(-Y_3/n))^n \\
& & \hspace*{0.6in}
 +
\exp(-t/n) \exp(-Y_1/n) \exp(-Y_2/n) \exp(-Y_3/n).
\end{eqnarray*}
Phrased more simply, if we define $Z_i = \exp(-Y_i/n)$, then the (2,2)
mirror theory is, as expected,
a ${\mathbb Z}_n$ orbifold with superpotential
\begin{displaymath}
W \: = \: - \sum_i \tilde{m}_i  n \ln Z_i +
Z_1^n + Z_2^n + Z_3^n + \exp(-t/n) Z_1 Z_2 Z_3,
\end{displaymath}
with the understanding that the fundamental fields are $Y_i$s not
$Z_i$s.  (For hypersurfaces, the fundamental fields will change.)

Later, we will use the matrices $(V^A_i)$ extensively, so in that language,
the change of variables above is encoded in
\begin{displaymath}
(V^A_i) \: = \: \left[ \begin{array}{cccc}
 1 & 0 & 0 & 1/n \\
 0 & 1 & 0 & 1/n \\
0 & 0 & 1 & 1/n \end{array} \right].
\end{displaymath}
Then, we write $Y_i = V^A_i \theta_A$, and so
\begin{displaymath}
Y_1 = \theta_1, \: \: \:
Y_2 = \theta_2, \: \: \:
Y_3 = \theta_3, \: \: \:
Y_p = (1/n)(\theta_1 + \theta_2 + \theta_3 - t).
\end{displaymath}

Let us next discuss the operator mirror map.
This is given by
\begin{eqnarray*}
\exp(-Y_1) = Z_1^n & \leftrightarrow & \sigma, \\
\exp(-Y_2) = Z_2^n & \leftrightarrow & \sigma, \\
\exp(-Y_3) = Z_3^n & \leftrightarrow & \sigma, \\
\exp(-Y_p) = Z_1 Z_2 Z_3 \exp(-t/n) & \leftrightarrow & -n \sigma.
\end{eqnarray*}

\subsection{(2,2) in (0,2) language}
\label{sect:22as02}

Now, let us describe (2,2) mirrors in (0,2) language, as preparation for
describing more general (0,2) mirrors.
Let $(\Sigma_a, \Upsilon_a)$ be the (0,2) chiral and Fermi components of
$\Sigma_a$, and $(Y_i, F_i)$ the (0,2) chiral and Fermi components of $Y_i$.
Then, the (2,2) superpotential~(\ref{eq:full22mirrorw}) is given in (0,2)
superspace by
\begin{equation} \label{eq:22w-02}
W \: = \: \sum_{a=1}^k \left[ \Upsilon_a \left( \sum_{i=1}^N Q^a_i Y_i
 - t_a\right)
\: + \: \sum_{i=1}^N \Sigma_a Q^a_i F_i \right]
 \: - \: \mu \sum_{i=1}^N F_i \exp(-Y_i) \: + \:
\sum_{i=1}^N \tilde{m}_i F_i.
\end{equation}
We integrate out $\Sigma_a$, $\Upsilon_a$ to get the constraints
\begin{displaymath}
\sum_{i=1}^N Q_i^a Y_i \: = \: t_a, \: \: \:
\sum_{i=1}^N Q_i^a F_i \: = \: 0,
\end{displaymath}
which we solve with the $V_i^A$ by writing
\begin{displaymath}
Y_i \: = \: \sum_{A=1}^{N-k} V_i^A \theta_A \: + \: \tilde{t}_i, \: \: \:
F_i \: = \: \sum_{A=1}^{N-k} V_i^A G_A,
\end{displaymath}
where $(\theta_A, G_A)$ are the chiral and Fermi components of the (2,2)
chiral superfields $\theta_A$.
After integrating out the constraints, the (0,2) superpotential becomes
\begin{equation}
W \: = \:  \sum_{i=1}^N \sum_{A=1}^{N-k} G_A V_i^A \left( \tilde{m}_i - 
\mu \exp(-Y_i) \right)
\: = \: 
\sum_{i=1}^N \sum_{A=1}^{N-k} G_A V_i^A \left( \tilde{m}_i - 
\mu e^{\tilde{t}_i} \prod_{B=1}^{N-k}
\exp(-V_i^B \theta_B) \right).
\end{equation}
As is standard, we remind that reader that
depending upon the entries in $(V^A_i)$,
the mirror may be a LG orbifold, which are required to leave
$W$ invariant.

In this language, the (2,2) mirror map between A- and B-model operators is
(partially) defined by
\begin{equation}
\sum_{a=1}^k Q_i^a \sigma_a + \tilde{m}_i \: \leftrightarrow \: \mu \exp(-Y_i)
\: = \: \mu e^{\tilde{t}_i} \prod_{A=1}^{N-k} \exp(-V_i^A \theta_A),
\end{equation}
which can be derived by differentiating~(\ref{eq:22w-02}) with respect to $F_i$.

In most of the rest of this paper, we will absorb $\mu$ into a field
redefinition of the $Y_i$s for simplicity, but we include it here
for completeness.

\section{Proposal for (0,2) Fano mirrors}
\label{sect:prop02}

We restrict to (0,2) theories obtained by (some) toric deformations of abelian
(2,2) GLSMs for Fano spaces, 
by which we mean physically that we choose $E$'s such that
$E_i \propto \phi_i$, where on the (2,2) locus $\phi_i$ is the chiral
superfield paired with the Fermi superfield whose superderivative is $E_i$.

In addition, to define a mirror, we also make another choice, namely 
we pick an invertible\footnote{
We assume that the charge matrix does indeed have an invertible $k \times k$
submatrix.  If not, then the theory has at least one free decoupled
$U(1)$, and after performing a change of basis to explicitly decouple
those $U(1)$'s, our analysis can proceed on the remainder.
} $k\times k$ submatrix, %which we will denote $S$,
of the charge matrix
$(Q_i^a)$, which we will denote $S$.  
The choice of $S$ will further constrain the allowed toric
deformations -- for a given $S$, we only consider some toric deformations.
Our mirror will depend upon the choice of $S$, and since different
$S$'s will yield different allowed bundle deformations, there need not
be a simple coordinate transformation relating results for different
choices of $S$ in general.
Furthermore, $S$ is only relevant for bundle deformations -- it does not
enter (2,2) locus computations, and so it has no analogue within
\cite{Hori:2000kt}.

For a given choice of $S$,
in the A/2 model, write 
\begin{displaymath}
E_i \: = \: \sum_{a=1}^k \sum_{j=1}^N \left(\delta_{ij} + B_{ij} \right)
Q_j^a \sigma_a \phi_i,
\end{displaymath}
where in the expression above, we do not sum over $i$'s.
The (0,2) deformations we will consider
are encoded in the matrices $B_{ij}$, where $B_{ij} = 0$
if $i$ defines a column of the matrix $S$.
Note that, at least on its face, this does not describe all possible
Euler-sequence-type (0,2) deformations, but only a special subset.
We will give a mirror construction for that special subset.

Then, 
the mirror can be described by
a collection of ${\mathbb C}^{\times}$-valued fields $Y_i$ 
(just as on the (2,2) locus, dual to the chiral superfields
of the original theory), satisfying the same D-term constraints as on the
(2,2) locus, and with (0,2) superpotential
\begin{eqnarray}
W & = &
\sum_{a=1}^k \left[ \Upsilon_a \left( \sum_{i=1}^N Q^a_i Y_i
 - t_a\right)
\: + \: \sum_{i=1}^N \Sigma_a Q^a_i F_i \right]
\nonumber \\
& &  - \mu \sum_i F_i \exp(-Y_i) \: + \: \mu \sum_{i} F_{i}
\left(
\sum_{i_S,j,a} B_{i j} Q_j^a [ (S^{-1})^T ]_{a i_S} \exp(-Y_{i_S})
\right),    \label{eq:02mirror-w-orig}
\end{eqnarray}
where $i_S$ denotes an index running through the columns of $S$,
and where the second term was chosen so that the resulting equations of
motion duplicate the chiral ring.
(For the moment, we have assumed no twisted masses are present; we will
return to twisted masses at the end of this section.)

Now, to do meaningful computations, we must apply the D-term constraints
to both $Y_i$'s and $F_i$'s.  Applying the D-term constraints to the $F_i$'s
to write them in terms of $G_A$'s ({\it i.e.} integrating out
$\Sigma_a$'s), and for simplicity
suppressing the $\Upsilon_a$ constraints and setting the mass scale
$\mu$ to unity, we have the expression
\begin{equation}  \label{eq:02mirror-w}
W \: = \: - \sum_{A=1}^{N-k} G_A \left( \sum_i V_i^A \exp(-Y_i) \: + \:
\sum_{i_S} D_{i_S}^A \exp(-Y_{i_S}) \right),
\end{equation}
where
\begin{equation}  \label{eq:d-defn}
D_{i_S}^A \: = \: - \sum_{i,j} \sum_a V_i^A B_{ij} Q_j^a [(S^{-1})^T]_{a i_S}
\end{equation}
Note when $B=0$, $D=0$, and the expression for $W$ above immediately reduces
to its (2,2) locus form.
We will derive this expression for $D$ below.

In this language, the mirror map between A/2- and B/2-model observables
is defined by
\begin{equation}    \label{eq:02mirror}
\sum_{a=1}^k \sum_{j=1}^N \left( \delta_{ij} + B_{ij} \right)
Q_j^a \sigma_a \: \leftrightarrow \:
\exp(-Y_i) \: = \:
 e^{\tilde{t}_i} \prod_{A=1}^{N-k} \exp(-V_i^A \theta_A).
\end{equation}
(Strictly speaking, we will see in examples that these equations define
not only the operator mirror map plus some of the chiral ring relations.)

We can derive the operator mirror map above from the 
superpotential~(\ref{eq:02mirror-w-orig}) by taking a derivative with
respect to $F_i$, as before.  Doing so, one finds
\begin{displaymath}
Q_i^a \sigma_a - \exp(-Y_i) + \sum_{i_S, j, a} B_{ij} Q_j^a [ (S^{-1} )^T
]_{a i_S} \exp(-Y_{i_S}) \: = \: 0.
\end{displaymath}
For $i$ corresponding to columns of $S$, $B_{ij} = 0$, and the expression
above simplifies to
\begin{displaymath}
S_{i_S}^a \sigma_a \: = \: \exp(-Y_{i_S}).
\end{displaymath}
Plugging this back in, we find
\begin{displaymath}
Q_i^a \sigma_a - \exp(-Y_i) + \sum_{j,a} B_{ij} Q_j^a \sigma_a \: = \: 0,
\end{displaymath}
which is easily seen to be the operator mirror map~(\ref{eq:02mirror}).

We can apply the operator mirror map
as follows.  Recall that the constraints imply
\begin{displaymath}
\sum_i Q_i^a Y_i \: = \: t_a
\end{displaymath}
hence
\begin{displaymath}
\prod_i \exp(-Q_i^a Y_i) \: = \: \exp(-t_a) \: = \: q_a,
\end{displaymath}
hence plugging in the proposed map~(\ref{eq:02mirror}) above, we have
\begin{displaymath}
\prod_i \left(
\sum_{a=1}^k \sum_{j=1}^N (\delta_{ij} + B_{ij} ) Q_j^a \sigma_a \right)^{Q_i^a}
\: = \: q_a,
\end{displaymath}
which is the chiral ring relation in the A/2-twisted GLSM.

In passing, to make the method above work, it is important that the
determinants appearing in quantum sheaf cohomology relations
in {\it e.g.} \cite{McOrist:2008ji,Donagi:2011uz,Donagi:2011va,Closset:2015ohf}
all factorize.  In other words, recall that for a general tangent bundle
deformation, the quantum sheaf cohomology ring relations take the form
\begin{displaymath}
\prod_{\alpha} ( \det M_{\alpha} )^{Q^a_{\alpha}} \: = \: q_a,
\end{displaymath}
where $\alpha$ denotes a block of chiral fields with the same
charges, and $M_{\alpha}$ encodes the $E$'s, which will mix chiral
superfields of the same charges.  In order for the operator 
mirror map construction
we have outlined above to work, it is necessary that each
$\det M_{\alpha}$ factorize into a product of factors, one for each
matter chiral multiplet.  This is ultimately the reason why in this paper we
have chosen to focus on `toric' deformations, in which each
$E$'s do not mix different matter chiral multiplets.

Now, in terms of the operator mirror map, let us derive the form of $D$ above
in equation~(\ref{eq:d-defn}).  The equations of motion from the
superpotential~(\ref{eq:02mirror-w}) are given by
\begin{displaymath}
\frac{\partial W}{\partial G_A} \: = \: 
 \sum_i V_i^A \exp(-Y_i) \: + \:
\sum_{i_S} D_{i_S}^A \exp(-Y_{i_S}) \: = \: 0.
\end{displaymath}
Now, we plug in the operator mirror map~(\ref{eq:02mirror}) above to get
\begin{displaymath}
\sum_i V_i^A \left( \sum_a \sum_j (\delta_{ij} + B_{ij} ) Q_j^a \sigma_a \right)
\: + \:
\sum_{i_S} D_{i_S}^A \left(
\sum_a \sum_j (\delta_{i_S j} + B_{i_S j} ) Q_j^a \sigma_a \right) \: = \: 0.
\end{displaymath}
Using the constraint
\begin{displaymath}
\sum_i V_i^A Q_i^a \: = \: 0,
\end{displaymath}
the first $\delta_{ij}$ term vanishes, and furthermore, since the matrix
$B$ is defined to vanish for indices from columns of $S$, we see that
in the second term, $B_{i_S j} = 0$, hence the equation above reduces to
\begin{displaymath}
\sum_{i, j} \sum_a V_i^A B_{ij} Q_j^a \sigma_a \: + \:
\sum_{i_S} \sum_a D_{i_S}^A S_{i_S}^a \sigma_a \: = \: 0.
\end{displaymath}
Since this should hold for all $\sigma_a$, we have that
\begin{displaymath}
\sum_{i, j} V_i^A B_{ij} Q_j^a \: + \: \sum_{i_S} D_{i_S}^A S^a_{i_S} \: = \: 0,
\end{displaymath}
which can be solved to give expression~(\ref{eq:d-defn}) for $D$ above.

Thus, the expression for the superpotential~(\ref{eq:02mirror-w}) 
together with the
operator 
mirror map~(\ref{eq:02mirror}) has equations of motion that duplicate the 
chiral ring.

In passing, one could also formally try to consider more general cases in
which a submatrix $S \subset Q$ is not specified.  One might then try to
take the expression for the mirror superpotential to be of the form
\begin{displaymath}
W \: = \: - \sum_{A=1}^{N-k} G_A \left(
\sum_i V_i^A \exp(-Y_i) + \sum_i D^A_i \exp(-Y_i) \right),
\end{displaymath}
where now the $i$ index on $D$ is allowed to run over all chiral superfields,
not just a subset.  Following the methods above, one cannot uniquely
solve for $D$ -- one gets families of possible $D$'s with undetermined
coefficients, and we do not know how to argue that the correlation functions
match for all such coefficients without restricting to subsets defined
by choices $S \subset Q$.

Now, in principle, for (0,2) theories defined by deformations of the 
(2,2) locus, there is an analogue of twisted masses that one can add
to the theory.  In the (2,2) case, twisted masses corresponded to replacing
a vector multiplet by its vevs, so that only a residue of $\sigma$
survived.  In (0,2), by contrast,
the vector multiplet does not contain $\sigma$,
only the gauge field, gauginos, and auxiliary fields $D$,
so we can no longer interpret the twisted mass in terms of replacing
a vector multiplet with its vevs.

Instead, we can understand the analogue of a twisted mass in a (0,2)
theory corresponding to a deformation of the (2,2) locus in terms of 
additions to $E_i = \overline{D}_+ \Lambda^i$, for Fermi superfields
$\Lambda^i$.  In particular, the (2,2) vector multiplet's $\sigma$ field
enters GLSMs written in (0,2) superfields
as a factor in such $E$'s, so twisted masses enter
similarly, as terms of the form
\begin{displaymath}
E_i \: = \: \tilde{m}_i \phi_i
\end{displaymath}
(where as usual we are admitting the possibilty of several toric
symmetries, and simply giving each chiral superfield the possibility of
its own twisted mass).
Such terms are only possible if the (0,2) superpotential has
compatible $J$'s, meaning that in order for supersymmetry to hold,
one requires 
$E \cdot J = 0$, as usual.  This is a residue of the requirement in the
(2,2) theory that twisted masses arise from flavor symmetries.

We have already seen, in section~\ref{sect:22as02}, how (2,2) twisted
masses can be represented in the mirror, described in (0,2) superspace.
To describe their combination with $E$ deformations is straightforward.
Briefly, the (0,2) mirror superpotential takes the form
\begin{equation}
W \: = \: - \sum_{A=1}^{N-k} G_A \left( \sum_i V_i^A \exp(-Y_i) \: + \:
\sum_{i_S} D_{i_S}^A \exp(-Y_{i_S}) \right) \: + \:
\sum_{i=1}^N \sum_{A=1}^{N-k} G_A V_i^A \tilde{m}_i,
\end{equation}
with $(D^A_{i_S})$ defined as in~(\ref{eq:d-defn}),
and the operator mirror map has the form
\begin{equation}
\sum_{a=1}^k \sum_{j=1}^N \left( \delta_{ij} + B_{ij} \right)
Q_j^a \sigma_a + \tilde{m}_i \: \leftrightarrow \:
\exp(-Y_i) \: = \:
 e^{\tilde{t}_i} \prod_{A=1}^{N-k} \exp(-V_i^A \theta_A).
\end{equation}

\section{Correlation functions}
\label{sect:correlation-fns}

In this section, we will argue formally that correlation functions in our
proposed (0,2) mirrors match those of the original theory.  More precisely,
we will compare closed-string
correlation functions of A- or A/2-twisted GLSM $\sigma$'s to
corresponding correlation functions in B- or B/2-twisted Landau-Ginzburg
models.  (Often, the Landau-Ginzburg mirror will be an orbifold; we will
only compare against untwisted sector correlation functions in such
orbifolds.)  Our computations will focus on genus zero computations, but in
(2,2) cases, in principle can be generalized to any genus.
(See also \cite{Closset:2017vvl} for related work.)

Before doing so, let us first outline in what sense correlation functions
match.  There are two possibilities:
\begin{itemize}
\item First, for special matrices $(V^A_i)$, we will argue that
correlation functions match on the nose.  In order for this to happen,
we will need to require that the determinant of an invertible
$k \times k$ submatrix of the charge matrix $Q$, match (up to sign)
the determinant of a complementary\footnote{
`Complementary' in this case means that if the $k \times k$ matrix
is defined by $i$'s corresponding to certain chiral superfields,
then those same chiral superfields cannot appear corresponding to
any $i$'s in the $(N-k)\times (N-k)$ submatrix of $(V^A_i)$.
} $(N-k) \times (N-k)$ submatrix
of $(V^A_i)$.
\item Alternatively, we can always formally rescale some of the $Y_i$s 
(without introducing or removing orbifolds) to arrange for the determinants
above to match, up to sign.  In this case, the correlation functions of
one theory are isomorphic to those of the other theory, but the
numerical factors will not match on the nose.  (Instead, the relations
between numerical factors will be determined by the rescaling of the
$Y_i$s.)
\end{itemize}
In either event, correlation functions will match.

\subsection{(2,2) supersymmetric cases}

We will first check that on the (2,2) locus, the ansatz described
above ({\it i.e.} the ansatz of \cite{Hori:2000kt}) generates
matching correlation functions between the A-twisted GLSM and
its B-twisted Landau-Ginzburg model mirror.
(See also \cite{Gomis:2012wy} 
for an analogous comparison of partition functions.)

First, let us consider correlation functions in an A-twisted GLSM.
An exact expression is given for fully massive cases in {\it e.g.}
\cite{Closset:2015rna}[equ'n (4.77)]:
\begin{displaymath}
\langle {\cal O} \rangle \: = \: \frac{ (-)^{N_c} }{|W|}
\frac{1}{(-2\pi i)^{{\rm rk}\, G} }
\sum_{\sigma_P} {\cal O} \, \frac{ Z_{\rm 1-loop}  }{H}
\end{displaymath}
where $G$ is the GLSM gauge group, $W$ its Weyl group, $N_c$ its rank,
\begin{displaymath}
Z_{\rm 1-loop} \: = \: \prod_{i=1}^N 
\left( \sum_{a=1}^k Q_i^a \sigma_a \right)^{
R(\Phi_i) - 1}, 
\end{displaymath}
($R(\Phi_i)$ the R-charge, which for simplicity we will assume to vanish,)
$\sigma_P$ the vacua,
and $H$ is the Hessian of the
twisted one-loop effective action, meaning
\begin{equation}   \label{eq:22amodel-hessian}
H \: = \: \det\left( \sum_i \frac{ Q^a_i Q^b_i }{\sum_c Q^c_i \sigma_c} \right),
\end{equation}
using (up to factors) the twisted one-loop effective action in
{\it e.g.} \cite{Morrison:1994fr}[equ'n (3.36)].

Now, up to irrelevant overall factors, 
there is an essentially identical expression for Landau-Ginzburg
correlation functions \cite{Vafa:1990mu}, 
involving the Hessian of the superpotential
rather than $H/Z_{\rm 1-loop}$ above.  
Therefore, to show that correlation functions match,
we will argue that the $H/Z_{\rm 1-loop}$ above, computed for the
A-twisted GLSM, matches the Hessian of superpotential derivatives for
the mirror Landau-Ginzburg model.  

First, since we are only interested in the determinant, we can rotate
the charge matrix $(Q^a_i)$ by an element $U \in SL(k,{\mathbb C})$
without changing the determinant:
\begin{displaymath}
\det\left( \sum_i \frac{ Q^a_i Q^b_i }{\sum_c Q^c_i \sigma_c} \right)
\mapsto 
(\det U)^2 \det\left( \sum_i \frac{ Q^a_i Q^b_i }{\sum_c Q^c_i \sigma_c} \right)
\: = \: \det\left( \sum_i \frac{ Q^a_i Q^b_i }{\sum_c Q^c_i \sigma_c} \right).
\end{displaymath}
Thus, it will be convenient to rotate the charge matrix to the form\footnote{
As we are not conjugating the charge matrix, but rather multiplying on one
side only, it should be possible to arrange for a $k \times k$ submatrix
to be diagonal, not just in Jordan normal form.
}
\be \label{QM}
Q^{a}_{i}=\left(
          \begin{array}{cccccc}
            a_{1} &  &  & Q^{1}_{k+1} & \cdots & Q^{1}_{N} \\
             & \ddots &  & \vdots & \ddots & \vdots \\
             &  & a_{k} & Q^{k}_{k+1} & \cdots & Q^{k}_{N}
          \end{array}
          \right).
\ee
Note that for the charge matrix in this form,
\begin{displaymath}
Z_{\rm 1-loop} \: = \: \left( \prod_{i=1}^k a_i \sigma_i \right)^{-1}
\left( \prod_{i=k+1}^N \left( \sum_{a=1}^k Q_i^a \sigma_a \right) \right)^{-1}.
\end{displaymath}
To put the charge matrix in this form, write
\begin{displaymath}
Q \: = \: [ S | * ] \: = \: (\det S)^{1/k} [ S' | * ],
\end{displaymath}
where $S$ is $k \times k$.  Then, multiply in $(S')^{-1}$, to get
\begin{displaymath}
(S')^{-1} Q \: = \: (\det S)^{1/k} [ I | * ],
\end{displaymath}
which is now diagonal.

It is straightforward to compute
\begin{eqnarray} %\label{QM3}
\lefteqn{
H = \det\left(\sum_{i}\frac{Q^{a}_{i}Q^{b}_{i}}{\sum_{c}Q^{c}_{i}\sigma_{c}}\right)
=
} \nonumber \\
& & \det\left(\begin{array}{ccc}
\frac{a_{1}}{\sigma_{1}}+\frac{(Q^{1}_{k+1})^{2}}{Q^{c}_{k+1}\sigma_{c}}+\ldots+\frac{(Q^{1}_{N})^{2}}{Q^{c}_{N}\sigma_{c}}  
& \frac{Q^{1}_{k+1}Q^{2}_{k+1}}{Q^{c}_{k+1}\sigma_{c}}+\ldots+\frac{Q^{1}_{N}Q^{2}_{N}}{Q^{c}_{N}\sigma_{c}}  
& \cdots \\
\frac{Q^{2}_{k+1}Q^{1}_{k+1}}{Q^{c}_{k+1}\sigma_{c}}+\ldots\frac{Q^{2}_{N}Q^{1}_{N}}{Q^{c}_{N}\sigma_{c}}  
& \frac{a_{2}}{\sigma_{2}}+\frac{(Q^{2}_{k+1})^{2}}{Q^{c}_{k+1}\sigma_{c}}+\ldots+\frac{(Q^{2}_{N})^{2}}{Q^{c}_{N}\sigma_{c}} 
& \cdots \\
            \vdots & \vdots & \ddots
          \end{array}
    \right).    \label{QM3}
\end{eqnarray}
We define\footnote{
The reader should note that the $t_i$ in this section, defined above,
is not related to $t$'s used earlier to describe FI parameters.
} $t_{i}=a_{i}/\sigma_{i}$ and $E^{a}_{i}=Q^{a}_{i}/\sqrt{\sum_{c}Q^{c}_{i}\sigma_{c}}$, then the matrix in the above determinant becomes
\begin{equation} \label{FMOA}
\left(
\begin{array}{ccc}
  t_{1} +(E^{1}_{k+1})^{2}+\ldots+(E^{1}_{N})^{2} & E^{1}_{k+1}E^{2}_{k+1}+\ldots+E^{1}_{N}E^{2}_{N} & \cdots \\
   E^{2}_{k+1}E^{1}_{k+1}+\ldots+E^{2}_{N}E^{1}_{N} & t_{2}+(E^{2}_{k+1})^{2}+\ldots+(E^{2}_{N})^{2} & \cdots \\
  \vdots & \vdots & \ddots
\end{array}
\right).
\end{equation}
When all the $t_{i}$ vanish, one can straightforwardly see that the matrix 
above is the product $(E^{T})^{T}E^{T}$, for matrix $E$
\begin{equation} \label{EM}
E=\left(
\begin{array}{cccc}
  E^{1}_{k+1} & E^{1}_{k+2} & \cdots & E^{1}_{N} \\
  E^{2}_{k+1} & E^{2}_{k+2} & \cdots & E^{2}_{N}\\
  \vdots & \vdots & \ddots & \vdots \\
   E^{k}_{k+1} & \cdots & \cdots & E^{k}_{N}
\end{array}
  \right).
\end{equation}
Using standard results from linear algebra, 
the generalized characteristic polynomial of matrix~(\ref{FMOA}), 
in terms of the variables $t_{i}$, is given by
\begin{equation} \label{AQMF}
\sum^{k}_{m=0}\left(\sum_{a_{1}<\cdots< a_{m}}
t_{a_{1}}\cdots  t_{a_{m}}\det\left(M_{a_{1}\cdots a_{m}}\right)\right),
\end{equation}
where the matrix $M_{a_{1}\cdots a_{m}}$ denotes the submatrix of 
$M=(E^{T})^{T}E^{T}$ by omitting rows $a_{1}\cdots a_{m}$ and 
columns $a'_{1}\cdots a'_{m}$ 
(\ie\quad a principal minor of $M$ of size $k-m$). 
(In our conventions, the determinant vanishes if $M$ has no entries.)
Notice that $M=(E^{T})^{T}E^{T}$, 
so the determinant can be written more simply as
\begin{eqnarray}\label{AQMF2}
\lefteqn{
\det M \: + \: t_1 \cdots t_k \: + \:
} \\
& &
\sum^{k-1}_{m=1}\left(\sum_{a_{1}<\cdots<a_{m}}
t_{a_{1}}\cdots  t_{a_{m}}
\left(\sum_{i_{1}<\cdots<i_{N+m-2k}}\left(\det E_{a_{1}\cdots a_{m}, 
i_{1}\cdots i_{N+m-2k}}\right)^{2}\right)\right),
\nonumber
\end{eqnarray}
where $\det E_{a_{1}\cdots a_{m}, i_{i}\cdots i_{N+m-2k}}$ denotes 
the determinant of the submatrix of $E$ formed by omitting
rows $a_{1}\cdots a_{m}$ and columns $i_{1}\cdots i_{N+m-2k}$.
(We formally require it to be zero when $N+m-2k<0$.)

Finally, we divide by $Z_{\rm 1-loop}$ to get an expression for
$H/Z_{\rm 1-loop}$ where $H$ is the Hessian~(\ref{eq:22amodel-hessian}):
\begin{eqnarray} \label{AHFF}
\lefteqn{
\frac{ \det M \: + \: t_1 \cdots t_k }{Z_{\rm 1-loop} } \: + \:
} \\
& &
\sum^{k-1}_{m=1}\left(\sum_{a_{1}<\cdots<a_{m}}
(a_{a_{1}})^{2}\cdots  (a_{a_{m}})^{2}
\left( \prod_{i\notin\{a_{1},\cdots ,a_{m}\}}
\left( \sum_{a=1}^k Q^{a}_{i}\sigma_{a}
\right) \right)
B_{a_1\cdots a_m} \right), \nonumber
\end{eqnarray}
for
\begin{eqnarray}
B_{a_1\cdots a_m} & = &
\sum_{i_{1}<\cdots<i_{N+m-2k}}\left(\det 
E_{a_{1}\cdots a_{m}, i_{1}\cdots i_{N+m-2k}}\right)^{2},
\nonumber
\end{eqnarray}
where $\det M$ vanishes for $N<2k$.
For later use, note that for $N \leq 2k$ we can expand 
\begin{equation}   \label{eq:detm-Z}
\frac{\det M}{Z_{\rm 1-loop}} \: = \: \left(
\prod_{i=1}^k a_i \sigma_i \right) \left(
\prod_{i \not\in \{i_1, \cdots, i_k\} }
\left( \sum_{c=1}^k Q_{k+i}^c \sigma_c \right) ( A_{i_1, \cdots, i_k} )^2
\right),
\end{equation}
and the terms for $1 \leq m \leq k-1$ are given by
\begin{equation}   \label{eq:detm-m}
a_{a_1}^2 \cdots a_{a_m}^2 \left( \prod_{b \not\in \{a_1, \cdots, a_m\} }
a_b \sigma_b \right) \left(
\prod_{i \not\in \{i_{m+1},\cdots,i_k\}} \left( \sum_{c=1}^k Q^c_{k+i} 
\sigma_c\right) \left( 
A_{i_{m+1}, \cdots, i_k}
\right)^2 \right),
\end{equation}
where $A_{i_{m+1},\cdots,i_k}$ denotes the sum of determinants of
all $(k-m)\times (k-m)$ submatrices of the charge matrix $(Q^a_i)$ for
values of $i > k$.

Next, we need to compare the ratio $H/Z_{\rm 1-loop}$
above to the analogous Hessian
arising in the mirror B-twisted Landau-Ginzburg model.
Here, there is a nearly identical computation in which the Hessian
we just computed is replaced with
the determinant of second derivatives of the mirror
superpotential~(\ref{eq:22mirrorw-basic}):
\begin{eqnarray*}
\frac{\partial^2 W}{\partial \theta_A \partial \theta_B}
& = &
- \sum_{i=1}^N \left( e^{\tilde{t}_i} \left(
\prod_{C=1}^{N-k} \exp(-V_i^C \theta_C) \right) V_i^A V_i^B \right), 
\\
& = & - \sum_{i=1}^N \left( \left( \sum_{a=1}^k Q_i^a \sigma_a \right) V_i^A
V_i^B
\right),
\end{eqnarray*}
using the operator mirror map~(\ref{eq:22mirror-basic}).

Thus, we need to compute
\begin{displaymath}
\det\left( \sum_i  V^A_i V^B_i \sum_c Q^c_i \sigma_c \right),
\end{displaymath}
and compare to the ratio $H/Z_{\rm 1-loop}$ 
from the A model that we computed previously.
In principle, the argument here is very similar to the argument just given
for the determinant defined by charge matrices.  First, using the fact
that $V$ has rank $N-k$, inside the determinant we can rotate $V$ to the
more convenient form
\be \label{VM}
V^{A}_{i}=\left(
          \begin{array}{cccccc}
            V^{1}_{1} & \cdots & V^{1}_{k} & \lambda^{1} &  &  \\
            \vdots & \ddots & \vdots &  & \ddots & \\
            V^{k}_{1} & \cdots & V^{k}_{k} &  &  & \lambda^{(N-k)}
          \end{array}
          \right).
\ee
In fact, we can say more.
Given that the $V$ matrix was originally defined to satisfy
\begin{displaymath}
\sum_i Q^a_i V^A_i \: = \: 0,
\end{displaymath}
after the rotation above inside the determinant, the $V$ matrix should
in fact have the form
\be \label{VM2}
V^{A}_{i}=\left(
 \begin{array}{cccccc}
            -\frac{\lambda^{1}Q^{1}_{k+1}}{a_{1}} & \cdots & -\frac{\lambda^{1}Q^{k}_{k+1}}{a_{k}} & \lambda^{1} &  &  \\
            \vdots & \ddots & \vdots &  & \ddots & \\
            -\frac{\lambda^{(N-k)}Q^{1}_{N}}{a_{1}} & \cdots & -\frac{\lambda^{(N-k)}Q^{k}_{N}}{a_{k}} &  &  & \lambda^{(N-k)}
          \end{array}
          \right).
\ee

Then, using the more convenient form of $V$ above,
we find that we can write the matrix
\begin{eqnarray}
\lefteqn{
\left( \sum_i  V^A_i V^B_i \sum_c Q^c_i \sigma_c \right)
\: =} \label{eq:mat1} \\
&  &
%\label{BHME}
\left( \begin{array}{ccc}
                              (\lambda^{1})^{2}\left[\frac{(Q^{1}_{k+1})^{2}\sigma_{1}}{a_{1}}+\cdots+\frac{(Q^{k}_{k+1})^{2}\sigma_{k}}{a_{k}}+Q^{c}_{k+1}\sigma_{c}\right] & \lambda^{1}\lambda^{2}\left[\frac{Q^{1}_{k+1}Q^{1}_{k+2}\sigma_{1}}{a_{1}}+\cdots+\frac{Q^{k}_{k+1}Q^{k}_{k+2}\sigma_{k}}{a_{k}}\right] & \cdots \\
                              \lambda^{2}\lambda^{1}\left[\frac{Q^{1}_{k+1}Q^{1}_{k+2}\sigma_{1}}{a_{1}}+\cdots+\frac{Q^{k}_{k+1}Q^{k}_{k+2}\sigma_{k}}{a_{k}}\right] & (\lambda^{2})^{2}\left[\frac{(Q^{1}_{k+2})^{2}\sigma_{1}}{a_{1}}+\cdots+\frac{(Q^{k}_{k+2})^{2}\sigma_{k}}{a_{k}}+Q^{c}_{k+2}\sigma_{c}\right] & \cdots \\
                              \vdots & \vdots & \ddots
                            \end{array}
                                                                       \right).
\nonumber
\end{eqnarray}
Similarly, we define $s_{i}=(\lambda^{i})^{2}Q^{c}_{k+i}\sigma_{c}$ and 
$F^{a}_{i}=\lambda^{i}Q^{a}_{k+i}\sqrt{\sigma_{a}/a_{a}}$ 
(without summing over the index $a$).
Then, the matrix above %(\ref{BHME}) 
can be written as
\begin{equation} \label{BMMF}
\left(
\begin{array}{ccc}
   s_{1}+(F^{1}_{1})^{2}+\cdots+(F^{k}_{1})^{2} &  F^{1}_{1}F^{1}_{2}+\cdots+ F^{k}_{1}F^{k}_{2} & \cdots \\
  F^{1}_{2}F^{1}_{1} +\cdots+ F^{k}_{2}F^{k}_{1} & s_{2}+(F^{1}_{2})^{2}+\cdots+(F^{k}_{2})^{2} & \cdots \\
  \vdots & \vdots & \ddots
\end{array}
\right).
\end{equation}
When all $s_{i}$ vanish, one can observe that the matrix is the
product $F^{T}F$, for
\begin{equation} \label{FM}
F=\left(
\begin{array}{cccc}
  F^{1}_{1} & F^{1}_{2} & \cdots & F^{1}_{N-k} \\
  F^{2}_{1} & F^{2}_{2} & \cdots & F^{2}_{N-k}\\
  \vdots & \vdots & \ddots & \vdots\\
   F^{k}_{1}& F^{k}_{2}  & \cdots &F^{k}_{N-k}
\end{array}
  \right).
\end{equation}
By using the same technique we can show that the determinant 
of~(\ref{eq:mat1}) vanishes for $N > 2k$, and for $N \leq 2k$ is
\begin{eqnarray} \label{BMDFE}
\lefteqn{
\det(F^T F) \: + \: s_1 \cdots s_{N-k} \: + \:
} \\
& &
\sum^{N-k-1}_{n=1}\left(\sum_{i_{1}<\cdots<i_{n}}
(s_{i_{1}} s_{i_2}\cdots s_{i_{n}})
\left(\sum_{a_{1}<\cdots<a_{ 2k-N+n}}\left(\det 
F_{i_{1}\cdots i_{n}, a_{1}\cdots a_{2k-N+n}}\right)^{2}\right)\right)
\nonumber
\end{eqnarray}
For later use, note that
\begin{eqnarray*}
\det F^T F & =  &
\sum_{a_1 < \cdots < a_{2k-N}} (\det F_{a_1 \cdots a_{2k-N}} )^2,
\\
& = & \left( \prod_{A=1}^{N-k} (\lambda^A)^2 \right)
\left( \prod_{b=1}^k \frac{\sigma_b}{a_b} \right)
\left( \sum_{i_1 < \cdots < i_{k}} \left(
\sum_{a_1, \cdots, a_k} Q^{a_{1}}_{k+i_1} \cdots
Q^{a_{k}}_{k+i_{k}} \epsilon_{a_1 \cdots a_{k}} \right)^2 \right),
\end{eqnarray*}
where
$F_{a_1 \cdots a_{2k-N}}$ denotes the submatrix of $F^a_i$ formed by deleting
columns $a_1$ through $a_{2k-N}$.

Next, we plug $$s_{i_{j}}=(\lambda^{i_{j}})^{2}Q^{c}_{k+i_{j}}\sigma_{c}$$ 
into equation~(\ref{BMDFE}), and compare equation~(\ref{AHFF}).
First, note that we can expand
\begin{displaymath}
\frac{t_1 \cdots t_k}{Z_{\rm 1-loop}} \: = \: \left( \prod_{i=1}^k a_i^2
\right)\left( \prod_{j=1}^N \left( \sum_{a=1}^k Q_j^a \sigma_a \right) \right),
\end{displaymath}
which matches
\begin{displaymath}
s_1 \cdots s_{N-k} \: = \: \prod_{A=1}^{N-k} (\lambda^A)^2 
\left( \sum_{a=1}^k Q_{k+A}^a \sigma_a \right)
\end{displaymath}
so long as
\begin{equation} \label{CDE}
\prod_{A=1}^{N-k} \lambda^A
=\pm 
\prod_{i=1}^k a_i.
\end{equation}
Analogous results hold for other terms, as we now verify.
First we consider the case $N \geq 2k$.
The term $\det M / Z_{\rm 1-loop}$ in the previous determinant corresponds
to the term $n=N-2k$ in the expansion~(\ref{BMDFE}), which is given by
\begin{displaymath}
\left( \prod_{A=1}^{N-k} \lambda^A \right)^2
\prod_{a=1}^k \frac{\sigma_a}{a_a} \prod_{i \not\in \{i_1, \cdots, i_k\} }
\left( \sum_{c=1}^k Q^c_{k+i} \sigma_c \right)
\left( A_{i_1 \cdots i_k} \right)^2,
\end{displaymath}
for $A_{i_1 \cdots i_k}$ defined previously.  It is easy to verify that
this matches equation~(\ref{eq:detm-Z}) for $\det M/Z_{\rm 1-loop}$ so
long as condition~(\ref{CDE}) is satisfied, just as before.
The remaining terms in expansion~(\ref{BMDFE}) for any given $n$
correspond to terms in~(\ref{AHFF}) with $m$ related by
$n = N-2k+m$, and have the form
\begin{displaymath}
\left( \prod_{A=1}^{N-k} \lambda^A \right)^2 \prod_{b \not\in \{
a_1, \cdots, a_m \} } \frac{\sigma_b}{a_b} \left(
\prod_{i \not\in \{i_{m+1}, \cdots, i_k\} }
\left( \sum_{c=1}^k Q_{k+i}^c \sigma_c \right)
\left( A_{i_{m+1} \cdots i_k} \right)^2 \right),
\end{displaymath}
and it is easy to verify that this matches equation~(\ref{eq:detm-m})
so long as condition~(\ref{CDE}) is satisfied, just as before.
The reader can now straightforwardly verify that analogous statements hold
for the cases $k < N < 2k$, which exhausts all nontrivial possibilities.

Thus, we see that correlation functions will match so long as
equation~(\ref{CDE}) holds.  Furthermore,
we can always arrange for equation~(\ref{CDE}) to hold.  If it does not do so
initially,
then as discussed at the start of this
section, we can perform field redefinitions and
rescale $Y_i$s to arrange for it to hold, at the cost
of making the isomorphism between the correlation functions of either
theory a shade more complicated.
For example, the coefficient of 
$$\left(\sum_{c}Q^{c}_{k+1}\sigma_{c}\right)\cdots 
\left(\sum_{d}Q^{d}_{N}\sigma_{d}\right)$$ 
in equation~(\ref{AHFF}) is $(a_{1}a_{2}\cdots a_{k})^{2}$, 
and the coefficient of the term of the same order in 
equation~(\ref{BMDFE}) is $(\lambda^{1} \lambda^{2}\cdots\lambda^{N-k})^{2}$. 
We see that equation~(\ref{CDE}) is required for equality hold, 
and the choice of sign in equation~(\ref{CDE}) should not have any physical 
significance.

So far, we have worked at genus zero, but the same argument also implies
that the same closed-string correlation functions match at arbitrary genus.
At genus $g$, A-twisted GLSM correlation functions are computed in the
same fashion albeit with a factor of $(H/Z_{\rm 1-loop})^{g-1}$
(see {\it e.g.} \cite{Nekrasov:2014xaa}[section 4], 
\cite{Benini:2015noa}[section 5.1]), whereas B-twisted Landau-Ginzburg
model correlation functions (in the untwisted sector) are computed with
a factor of $(H')^{g-1}$ \cite{Vafa:1990mu}, for $H'$ the determinant of second
derivatives of the mirror superpotential.  Demonstrating that
$H/Z_{\rm 1-loop} = H'$ therefore not only demonstrates that genus
zero correlation functions match, but also higher-genus correlation
functions.  (For (0,2) theories, by contrast, higher genus correlation
functions are not yet understood, so there we will only be able to
compare genus zero correlation functions.)

Essentially the same argument applies if one adds twisted masses to the
theory.  One simply makes the substitution
\begin{equation} \label{TM}
\sum_{a}Q^{a}_{i}\sigma_{a}\rightarrow \sum_{a}Q^{a}_{i}\sigma_{a}+\widetilde{m}_{i},
\end{equation}
where $\widetilde{m}_{i}$ is the twisted mass.  The details of the proof
above are essentially unchanged.  Also note that we are free to redefine
$\sigma_{a}$ to $\sigma_{a}+c_{a}$, 
and we can use this to set the first $k$ twisted masses to zero. 
This leaves $N-k$ twisted masses, consistent with a global
flavor symmetry $U(1)^{N-k}$.

The arguments above hold so long as one can integrate out all of the
matter Higgs fields, to obtain a pure Coulomb branch.  In the (2,2) theory
one expects that one should be able to do this if one adds sufficient
twisted masses (see {\it e.g.} \cite{Nekrasov:2014xaa}[section 2.3]).
(In particular, adding twisted masses can act as a substitute for going
far out along the Coulomb branch, which also makes the matter fields 
massive.)

\subsection{(0,2) supersymmetric cases}

We will now generalize the previous argument to (0,2) cases.

Our argument here will be very similar to that given for (2,2) cases.
We will compare results for correlation functions in A/2-twisted
GLSMs computed with supersymmetric localization to results for
correlation functions computed in B/2-twisted (0,2) Landau-Ginzburg models.

First, as before, applying supersymmetric localization to an A/2-twisted
GLSM, there is an exact formula for (genus zero) (0,2) correlation functions
\cite{Closset:2015ohf},
which has more or less the same form as in the (2,2) case, now involving
a Hessian of derivatives of a twisted one-loop (0,2) effective action
\cite{McOrist:2008ji}, which takes the form
\begin{equation}  \label{eq:02-det-a}
H \: = \: \det\left( \sum_i \frac{ \sum_j Q^a_i A_{ij} Q^b_j }{ 
\sum_m A_{i m} Q^c_m \sigma_c } \right),
\end{equation}
where $A_{ij} = \delta_{ij} + B_{ij}$.

We assume without loss of generality that the invertible
$S$ submatrix of the charge matrix corresponds to the first $k$ columns
of $Q$.  Then, one can show that the determinant~(\ref{eq:02-det-a}) above
is equal to
\begin{equation} \label{AHE02}
\det\left(\begin{array}{ccc}
\frac{a_{1}}{\sigma_{1}}+\frac{Q^{1}_{k+1}\left(Q^{1}_{k+1}+\varepsilon^{1}_{k+1}\right)}{\left(Q^{a}_{k+1}+\varepsilon^{a}_{k+1}\right)\sigma_{a}}+\ldots+\frac{Q^{1}_{N}\left(Q^{1}_{N}+\varepsilon^{1}_{N}\right)}{\left(Q^{a}_{N}+\varepsilon^{a}_{N}\right)\sigma_{a}} 
 & \frac{Q^{1}_{k+1}\left(Q^{2}_{k+1}+\varepsilon^{2}_{k+1}\right)}{\left(Q^{a}_{k+1}+\varepsilon^{a}_{k+1}\right)\sigma_{a}}+\ldots\frac{Q^{1}_{N}\left(Q^{2}_{N}+\varepsilon^{2}_{N}\right)}{\left(Q^{a}_{N}+\varepsilon^{a}_{N}\right)\sigma_{a}} 
 & \cdots \\
 \frac{Q^{2}_{k+1}\left(Q^{1}_{k+1}+\varepsilon^{1}_{k+1}\right)}{\left(Q^{a}_{k+1}+\varepsilon^{a}_{k+1}\right)\sigma_{a}}+\ldots\frac{Q^{2}_{N}\left(Q^{1}_{N}+\varepsilon^{1}_{N}\right)}{\left(Q^{a}_{N}+\varepsilon^{a}_{N}\right)\sigma_{a}} 
 & \frac{a_{2}}{\sigma_{2}}+\frac{Q^{2}_{k+1}\left(Q^{2}_{k+1}+\varepsilon^{2}_{k+1}\right)}{\left(Q^{a}_{k+1}+\varepsilon^{a}_{k+1}\right)\sigma_{a}}+\ldots+\frac{Q^{2}_{N}\left(Q^{2}_{N}+\varepsilon^{2}_{N}\right)}{(Q^{a}_{N}+\varepsilon^{a}_{N})\sigma_{a}}
 & \cdots \\
            \vdots & \vdots & \ddots
          \end{array}
    \right),
\end{equation}
where $\varepsilon^{a}_{i}=\sum_{j}B_{ij}Q^{a}_{j}$.

Now, in the B/2-twisted (0,2) Landau-Ginzburg model, there is
an analogous expression for correlation functions
\cite{Melnikov:2007xi}, involving the Hessian
\begin{displaymath}
\det \frac{\partial^2 W}{\partial G_A \partial \theta_B}.
\end{displaymath}
One can similarly show that the Hessian above
is given by (using the (0,2) operator mirror map~(\ref{eq:02mirror}))
\begin{equation} \label{BHE02}
\det\left( \begin{array}{ccc}
                              (\lambda^{1})^{2}\left[\sum^{k}_{b=1}\frac{Q^{b}_{k+1}\left(Q^{b}_{k+1}+\varepsilon^{b}_{k+1}\right)\sigma_{b}}{a_{b}}+S_{k+1}\right] & \lambda^{1}\lambda^{2}\left[\sum^{k}_{b=1}\frac{\left(Q^{b}_{k+1}+\varepsilon^{b}_{k+1}\right)Q^{b}_{k+2}\sigma_{b}}{a_{b}}\right] & \cdots \\
                              \lambda^{2}\lambda^{1}\left[\sum^{k}_{b=1}\frac{Q^{b}_{k+1}\left(Q^{b}_{k+2}+\varepsilon^{b}_{k+2}\right)\sigma_{b}}{a_{b}}\right] & (\lambda^{2})^{2}\left[\sum^{k}_{b=1}\frac{Q^{b}_{k+2}\left(Q^{b}_{k+2}+\varepsilon^{b}_{k+2}\right)\sigma_{b}}{a_{b}}+S_{k+2}\right] & \cdots \\
                              \vdots & \vdots & \ddots
                            \end{array}
                                                                       \right),
\end{equation}
where 
$S_{k+i}=\sum_{a}\left(Q^{a}_{k+i}+\varepsilon^{a}_{k+i}\right)\sigma_{a}$.

Finally, following the same steps as for (2,2),
one can show that the ratio $H/Z_{\rm 1-loop}$ appearing in the A/2-twisted
GLSM matches the Hessian appearing in the B/2-twisted Landau-Ginzburg model,
\begin{displaymath}
\det\left( \sum_i \frac{ \sum_j Q^a_i A_{ij} Q^b_j }{\sum_m A_{im} Q^c_m 
\sigma_c} \right) \left( \prod_{i=1}^k a_i \sigma_i \right)
\left( \prod_{j=k+1}^N ( Q^a_j + \epsilon^a_j)
\sigma_a\right) \: = \:
\det \frac{\partial^2 W}{\partial G_A \partial \theta_B},
\end{displaymath}
so long as
\begin{displaymath}
\prod_{i=1}^{N-k} \lambda^i \: = \: \pm \prod_{i=1}^k a_i .
\end{displaymath}
(As before, if this does not hold, we can always perform field redefinitions
to rescale some of the $Y_i$s and corresponding Fermi fields $F_i$, at the cost of making the isomorphism
between correlation functions of either theory slightly more
complicated.)
Thus, the A/2-twisted GLSM Hessian matches that arising in 
B/2-twisted Landau-Ginzburg model correlation functions
\cite{Melnikov:2007xi}.
Since correlation functions in the A/2-twisted GLSM and the B/2-twisted
(0,2) Landau-Ginzburg model have essentially the same form, albeit with
potential different Hessians, and we have now demonstrated that the
Hessians match, it follows that correlation functions match.

\section{Examples}
\label{sect:exs}

So far we have presented formal arguments for a (0,2) mirror 
defined by a (0,2) Landau-Ginzburg theory with the same chiral ring
and correlation functions\footnote{
Technically, untwisted sector correlation functions, if the mirror involves
an orbifold.
}
as the original A/2 theory.  In this section we will verify that this
proposal reproduces known results in specific examples.

To be specific, we will compare predictions to mirror proposals previously
made in \cite{Chen:2016tdd,Chen:2017mxp}.  
Those papers were originally written by guessing ansatzes for possible 
mirrors, constrained to match known results on the (2,2) locus and to have
the correct correlation functions and chiral ring relations.  Here,
we will see that the proposal we have presented correctly and systematically
reproduces the results obtained by much more laborious methods in
\cite{Chen:2016tdd,Chen:2017mxp}.  
This will implicitly also provide tests that correlation functions and
chiral rings do indeed match, as argued formally in the last section.

That said, the systematic proposal of this paper will only apply to
special, `toric' deformations, not all tangent bundle deformations,
not even all tangent bundle deformations realizable by Euler-type
sequences.  Curiously, the terms in the mirrors described in
\cite{Chen:2016tdd,Chen:2017mxp} that are not realized are nonlinear
in the fields, suggesting that toric deformations are mirror to
linear terms.  We will not pursue this direction further in this paper,
but mention it here for completeness.

\subsection{${\mathbb P}^1 \times {\mathbb P}^1$}

We begin by reviewing the now-nearly-canonical example of
${\mathbb P}^1 \times {\mathbb P}^1$.  The charge matrix of the 
GLSM for the chiral fields is of the form
\begin{displaymath}
\left[ \begin{array}{cccc}
1 & 1 & 0 & 0 \\
0 & 0 & 1 & 1
\end{array} \right],
\end{displaymath}
and deformations of the tangent bundle are described mathematically as the
cokernel of the short exact sequence
\begin{displaymath}
0 \: \longrightarrow \: {\cal O}^2 \: \stackrel{*}{\longrightarrow} \:
{\cal O}(1,0)^2 \oplus {\cal O}(0,1)^2 \: \longrightarrow \: 
{\cal E} \: \longrightarrow \: 0,
\end{displaymath}
where the map $*$ is given by
\begin{displaymath}
* \: = \: \left[ \begin{array}{cc}
\tilde{A} x & \tilde{B} x \\
\tilde{C} y & \tilde{D} y \end{array} \right],
\end{displaymath}
where $\tilde{A}$, $\tilde{B}$, $\tilde{C}$, $\tilde{D}$ 
are constant $2 \times 2$ matrices, and $x$, $y$
are two-component vectors of homogeneous coordinates on either 
${\mathbb P}^1$ factor.

We have the same constraints on fields from D terms as on the (2,2) locus,
namely:
\begin{displaymath}
Y_1 \: + \: Y_2 \: = \: t_1, \: \: \:
Y_3 \: + \: Y_4 \: = \: t_2,
\end{displaymath}
where $Y_{1,2}$ are dual to the variables for one ${\mathbb P}^1$,
and $Y_{3,4}$ are dual to the variables for the other.

Let us solve the constraints above by taking
\begin{displaymath}
\tilde{t}_1 \: = \: 0, \: \: \:
\tilde{t}_2 \: = \: t_1, \: \: \:
\tilde{t}_3 \: = \: 0, \: \: \:
\tilde{t}_4 \: = \: t_2,
\end{displaymath}
and
\begin{displaymath}
V_i^A \: = \: \left[ \begin{array}{cccc}
1 & -1 & 0 & 0 \\
0 & 0 & 1 & -1 \end{array} \right],
\end{displaymath}
so that
\begin{displaymath}
Y_1 \: = \: \theta, \: \: \: Y_2 \: = \: t_1 \: - \: \theta, \: \: \:
G^1 \: = \: F_1 \: = \: - F_2,
\end{displaymath}
\begin{displaymath}
Y_3 \: = \: \tilde{\theta}, \: \: \: Y_4 \: = \: t_2 \: - \: \tilde{\theta},
\: \: \:
G^2 \: = \:  F_3 \: = \: - F_4.
\end{displaymath}

\subsubsection{First choice of $S$}

We will first consider the case that
the submatrix $S$ is given by the first and third
columns of the charge matrix $Q$ above, so that
$S$ is the identity matrix.  Then, the allowed deformations are defined
by
\begin{displaymath}
A_{ij} \: = \: \left[ \begin{array}{cccc}
1 & 0 & 0 & 0 \\
A_{21} & A_{22}  & A_{23} & A_{24}  \\
0 & 0 & 1 & 0 \\
A_{41} & A_{42}  & A_{43} &  A_{44}
\end{array} \right].
\end{displaymath}
(Rows correspond to fixed $i$ index, and reflect the fact that for
$i$ a row of $S$, values are fixed to those of the (2,2) locus.)
We can find the corresponding tangent bundle deformations by comparing the
$E_i$'s.  In terms of the matrix above,
\begin{eqnarray*}
E_1 & = & \sum_{j,a} A_{1j} Q_j^a \sigma_a \phi_1, \\
& = & \sigma_1 \phi_1, \\
E_2 & = & \left( (A_{21}+A_{22}) \sigma_1 + (A_{23}+A_{24}) \sigma_2 
\right) \phi_2, \\
E_3 & = & \sigma_2 \phi_3, \\
E_4 & = & \left( A_{41}+A_{42}) \sigma_1 + (A_{43}+A_{44}) \sigma_2
\right) \phi_4,
\end{eqnarray*}
whereas in terms of $\tilde{A}, \tilde{B}, \tilde{C}, \tilde{D}$,
we have
\begin{eqnarray*}
E_1 & = & (\tilde{A}_{11} \phi_1 + \tilde{A}_{12} \phi_2) \sigma_1 +
(\tilde{B}_{11} \phi_1 + \tilde{B}_{12} \phi_2 ) \sigma_2, \\
E_2 & = & (\tilde{A}_{21} \phi_1 + \tilde{A}_{22} \phi_2) \sigma_1 +
(\tilde{B}_{21} \phi_1 + \tilde{B}_{22} \phi_2) \sigma_2, \\
E_3 & = & (\tilde{C}_{11} \phi_3 + \tilde{C}_{12} \phi_4) \sigma_1 +
(\tilde{D}_{11} \phi_3 + \tilde{D}_{12} \phi_4) \sigma_2, \\
E_4 & = & (\tilde{C}_{21} \phi_3 + \tilde{C}_{22} \phi_4) \sigma_1 +
(\tilde{D}_{21} \phi_3 + \tilde{D}_{22} \phi_4) \sigma_2,
\end{eqnarray*}
for $\phi_{1,2}$ corresponding to homogeneous coordinates on the first
${\mathbb P}^1$ factor
and $\phi_{3,4}$ corresponding to homogeneous coordinates
on the second.  Comparing these two expressions, we find that
in terms of the original $2 \times 2$ matrices, our ansatz is equivalent
to the special case
\begin{displaymath}
\tilde{A} \: = \: \left[ \begin{array}{cc}
1 & 0 \\
0 & A_{21}+A_{22} \end{array} \right], \: \: \:
\tilde{B} \: = \: \left[ \begin{array}{cc}
0 & 0 \\
0 & A_{23}+A_{24} \end{array} \right],% \: \: \:
\end{displaymath}
\begin{displaymath}
\tilde{C} \: = \: \left[ \begin{array}{cc}
0 & 0 \\
0 & A_{41}+A_{42} \end{array} \right], \: \: \:
\tilde{D} \: = \: \left[ \begin{array}{cc}
1 & 0 \\
0 & A_{43}+A_{44} \end{array} \right].
\end{displaymath}

From formula~(\ref{eq:d-defn}), we have that
\begin{displaymath}
(D^A_{i_S}) \: = \: \left[ \begin{array}{cc}
A_{21}+A_{22}-1 &  A_{23}+A_{24} \\
A_{41}+A_{42} & A_{43}+A_{44}-1 \end{array} \right],
\end{displaymath}
where the $A$ index counts across rows, and the $i_S$ index counts
across columns.

The (0,2) superpotential of the proposed mirror is then given 
by~(\ref{eq:02mirror-w})
\begin{eqnarray*}
W & = & - G_1 \left( e^{-Y_1} - e^{-Y_2} + (A_{21}+A_{22}-1) e^{-Y_1} + 
(A_{23}+A_{24}) e^{-Y_3} \right)
\\
& & - G_2\left( e^{-Y_3} - e^{-Y_4} + (A_{41}+A_{42}) e^{-Y_1} + 
(A_{43}+A_{44}-1) e^{-Y_3} \right), \\
\end{eqnarray*}
If we define %$X_1 = \exp(\theta)$, $X_2 = \exp(\tilde{\theta})$,
$X_i = \exp(-Y_i)$,
then the (0,2) superpotential above can be written
\begin{eqnarray*}
W & = & - G_1 \left( (A_{21}+A_{22}) X_1 - \frac{q_1}{X_1} 
+ (A_{23}+A_{24}) X_3 \right) \\
& & - G_2\left( (A_{43}+A_{44}) X_3 - \frac{q_2}{X_3} + (A_{41}+A_{42}) X_1 
 \right).
\end{eqnarray*}
The operator mirror map~(\ref{eq:02mirror}) in this case implies
\begin{eqnarray*}
X_1 & \leftrightarrow & \sigma_1, \\
X_3 & \leftrightarrow & \sigma_2.
\end{eqnarray*}
In fact, the operator mirror map statement earlier also implies
\begin{eqnarray}
X_2 = \frac{q_1}{X_1} & \leftrightarrow & (A_{21} + A_{22}) \sigma_1 +
(A_{23}+A_{24}) \sigma_2, \label{eq:p1p1:ex1:m1}\\
X_4 = \frac{q_2}{X_3} & \leftrightarrow & (A_{41}+A_{42}) \sigma_1 +
(A_{43}+A_{44}) \sigma_2.   \label{eq:p1p1:ex1:m2}
\end{eqnarray}
These two statements are also redundant consequences of the
equations of motion $\partial W/\partial G_{1,2} = 0$, which
are themselves consequences
of the quantum sheaf cohomology
relations, as we shall see momentarily.

Now, let us compare results from \cite{Chen:2016tdd}.
There, it was argued that the (0,2) mirror could be represented as
\begin{eqnarray*}
W & = & - G_1 \left( a X'_1 + b \frac{(X'_2)^2}{X'_1} + \mu X'_2 - 
\frac{q_1}{X'_1} \right) \\
& & - G_2 \left( d X'_2 + c \frac{(X'_1)^2}{X'_2} + \nu X'_1 - \frac{q_2}{X'_2}
\right),
\end{eqnarray*}
where
\begin{displaymath}
a = \det \tilde{A}, \: \: \:
b = \det \tilde{B}, \: \: \:
c = \det \tilde{C}, \: \: \:
d = \det \tilde{D},
\end{displaymath}
\begin{eqnarray*}
\mu & = & \det(\tilde{A} + \tilde{B}) - \det \tilde{A} - \det \tilde{B}, \\
\nu & = & \det( \tilde{C} + \tilde{D}) - \det \tilde{C} - \det \tilde{D},
\end{eqnarray*}
and with operator mirror map
\begin{displaymath}
X'_1 \leftrightarrow \sigma_1, \: \: \:
X'_2 \leftrightarrow \sigma_2.
\end{displaymath}

In the present case, 
\begin{displaymath}
a = A_{21}+A_{22}, \: \: \:
b = 0, \: \: \:
c = 0, \: \: \:
d = A_{43}+A_{44}, \: \: \:
\mu = A_{23}+A_{24}, \: \: \:
\nu = A_{41}+A_{42},
\end{displaymath}
and it is easy to see that the mirror superpotential proposed here
matches the specialization of that in \cite{Chen:2016tdd},
if we identify $X_1 = X'_1$, $X_3 = X'_2$..

In addition, the quantum sheaf cohomology ring in this model is
\cite{McOrist:2008ji,Donagi:2011uz,Donagi:2011va,Closset:2015ohf}
\begin{displaymath}
a \sigma_1^2 + b\sigma_2 + \mu \sigma_1 \sigma_2 = q_1, \: \: \:
c \sigma_1^2 + d \sigma_2^2 + \nu \sigma_1 \sigma_2 = q_2,
\end{displaymath}
which in the present case matches the remaining mirror
map equations~(\ref{eq:p1p1:ex1:m1}), (\ref{eq:p1p1:ex1:m2}) above.

Altogether, we see that the mirror proposed here matches results in
\cite{Chen:2016tdd}, giving us a consistency check not only of proposed
mirror superpotentials but also implicitly correlation functions and
chiral rings.

\subsubsection{Second choice of $S$}

Next, we consider the case that
the submatrix $S$ is given by the first and fourth
columns of the charge matrix $Q$, so that $S$ is the identity matrix,
and the allowed deformations are
\begin{displaymath}
(A_{ij}) \: = \: \left[ \begin{array}{cccc}
1 & 0 & 0 & 0 \\
A_{21} & A_{22} & A_{23} & A_{24} \\
A_{31} & A_{32} & A_{33} & A_{34} \\
0 & 0 & 0 & 1
\end{array} \right].
\end{displaymath}
(Rows correspond to fixed $i$ index, and reflect the fact that for $i$
a row of $S$, values are fixed to those of the (2,2) locus.)
In other words, in terms of the original $2 \times 2$ matrices, we restrict
to the special case
\begin{displaymath}
\tilde{A} \: = \: \left[ \begin{array}{cc}
1 & 0 \\
0 & A_{21}+A_{22} \end{array} \right], \: \: \:
\tilde{B} \: = \: \left[ \begin{array}{cc}
0 & 0 \\
0 & A_{23}+A_{24} \end{array} \right], %\: \: \:
\end{displaymath}
\begin{displaymath}
\tilde{C} \: = \: \left[ \begin{array}{cc}
A_{31}+A_{32}& 0 \\
0 & 0 \end{array} \right], \: \: \:
\tilde{D} \: = \: \left[ \begin{array}{cc}
A_{33}+A_{34} & 0 \\
0 & 1 \end{array} \right].
\end{displaymath}

From formula~(\ref{eq:d-defn}), we have that
\begin{displaymath}
(D^A_{i_S}) \: = \: \left[ \begin{array}{cc}
A_{21}+A_{22}-1 & A_{23}+A_{24} \\
-(A_{31}+A_{32}) & -(A_{33}+A_{34}-1) \end{array} \right],
\end{displaymath}
where the $A$ index counts across rows, and the $i_S$ index counts
across columns.

The (0,2) superpotential of the proposed mirror~(\ref{eq:02mirror-w}) 
is then given by
\begin{eqnarray*}
W & = & - G_1 \left( e^{-Y_1} - e^{-Y_2} + (A_{21}+A_{22} - 1) e^{-Y_1} +
(A_{23}+A_{24}) e^{-Y_4} \right) 
\\
& & - G_2\left( e^{-Y_3} - e^{-Y_4} - (A_{31}+A_{32}) e^{-Y_1} - (A_{33}+A_{34}-1)
e^{-Y_4} \right), \\
& = & - G_1 \left( (A_{21}+A_{22}) X_1 - \frac{q_1}{X_1} + (A_{23}+A_{24})
X_4 \right) \\
& & - G_2 \left(\frac{q_2}{X_4} - (A_{31}+A_{32}) X_1 - (A_{33}+A_{34}) X_4 
 \right),
\end{eqnarray*}
where $X_i = \exp(-Y_i)$ and with hindsight we have chosen to write the
superpotential in terms of $X_1$ and $X_4$, to illuminate the relation to
other mirrors.

Furthermore, the operator mirror map~(\ref{eq:02mirror}) implies
\begin{eqnarray}
X_1 & \leftrightarrow & \sigma_1, \label{eq:p1p1:ex2:mm1}\\
X_4 & \leftrightarrow & \sigma_2
\label{eq:p1p1:ex2:mm2}
\end{eqnarray}
and also implies
\begin{eqnarray}
X_2 = \frac{q_1}{X_1} & \leftrightarrow & 
(A_{21}+A_{22}) \sigma_1 + (A_{23}+A_{24}) \sigma_2, \label{eq:p1p1:ex2:mm3} \\
X_3  = \frac{q_2}{X_4} & \leftrightarrow & (A_{31}+A_{32}) \sigma_1 +
(A_{33}+A_{34}) \sigma_2   
  \label{eq:p1p1:ex2:mm4}
\end{eqnarray}
The latter two relations are a redundant restatement of the chiral
ring of the theory, which can be seen by comparing the equations of
motion of the superpotential above, defined by $\partial W/\partial G_A = 0$.

Now, in this case the mirror given in \cite{Chen:2016tdd} is defined by
\begin{displaymath}
a = A_{21}+A_{22}, \: \: \:
b = c= 0, \: \: \:
d = A_{33}+A_{34}, \: \: \:
\mu = A_{23}+A_{24}, \: \: \:
\nu = A_{31}+A_{32},
\end{displaymath}
and so has the form
\begin{eqnarray*}
W & = & - G_1 \left( (A_{21}+A_{22}) X'_1 + (A_{23}+A_{24}) X'_2 - \frac{q_1}{X'_1}
\right)
\\
& & - G_2 \left( (A_{33}+A_{34})X'_2 +(A_{31}+A_{32})X'_1 - \frac{q_2}{X'_2} \right),
\end{eqnarray*}
with operator mirror map 
\begin{displaymath}
X'_1 \leftrightarrow \sigma_1, \: \: \:
X'_2 \leftrightarrow \sigma_2.
\end{displaymath}
With the dictionary $X_1 = X'_1$, $X_4 = X'_2$ and a sign change on $G_2$,
we see that the superpotential and operator mirror map predicted here match
that in \cite{Chen:2016tdd}.  In passing, note that
the fact that we reproduce the mirror of \cite{Chen:2016tdd} implicitly
gives an independent check of matching of correlation functions.

\subsubsection{Comment on Adams-Basu-Sethi result}

Let us very briefly compare to the analysis of \cite{Adams:2003zy},
which
also examined (0,2) mirrors to
${\mathbb P}^1 \times {\mathbb P}^1$.
They primarily considered tangent
bundle deformations of the form \cite{Adams:2003zy}[equ'n (255)]
\begin{displaymath}
\tilde{A} = I = \tilde{D}, \: \: \:
\tilde{C} = 0, \: \: \:
\tilde{B} = \left[ \begin{array}{cc}
\epsilon_1 & 0 \\
0 & \epsilon_2 \end{array} \right].
\end{displaymath}
In principle, a deformation of this form is compatible with making
a non-invertible choice of $S$, specifically the last two columns of
the charge matrix, so that
\begin{displaymath}
S \: = \: \left[ \begin{array}{cc}
0 & 0 \\
1 & 1 \end{array} \right].
\end{displaymath}
Since $S$ is not invertible, it is not possible to uniquely solve
for $(D^A_{i_S})$i in our approach, 
and we find it intriguing that in their analysis,
they also were not able to uniquely solve for the mirror superpotential
without doing further worldsheet instanton computations to solve for
the values of some otherwise-undetermined parameters.

\subsection{${\mathbb F}_n$}

Next we consider Hirzebruch surfaces ${\mathbb F}_n$.  Now,
for $n>1$, these are not Fano, but nevertheless one can write down a mirror
for the GLSM (which for the non-Fano cases is more properly interpreted
as a mirror to a different geometric phase, the UV phase,
of the GLSM), as discussed
in \cite{Chen:2017mxp}.  The charge matrix
of the GLSM is
\begin{displaymath}
\left[ \begin{array}{cccc}
1 & 1 & n & 0 \\
0 & 0 & 1 & 1 \end{array} \right],
\end{displaymath}
and deformations of the tangent bundle are described mathematically as the
cokernel ${\cal E}$ of the short exact sequence
\begin{displaymath}
0 \: \longrightarrow \: {\cal O}^2 \: \stackrel{*}{\longrightarrow} \:
{\cal O}(1,0)^2 \oplus {\cal O}(n,1) \oplus {\cal O}(0,1) \:
\longrightarrow \: {\cal E} \: \longrightarrow \: 0,
\end{displaymath}
where
\begin{displaymath}
* \: = \: \left[ \begin{array}{cc}
\tilde{A}x & \tilde{B}x \\
\gamma_1 s & \gamma_2 s \\
\alpha_1 t & \alpha_2 t \end{array} \right].
\end{displaymath}
In the expression above, $x$ is a two-component vector of homogeneous
coordinates of charge $(1,0)$, 
$s$ is a homogeneous coordinate of charge $(n,1)$,
and $t$ is a homogeneous coordinate of charge $(0,1)$,
$A$, $B$ are constant 
$2 \times 2$ matrices, and $\gamma_{1,2}$, $\alpha_{1,2}$ are
constants.
(In principle, nonlinear deformations are also possible, but as
observed previously in {\it e.g.} 
\cite{McOrist:2008ji,Donagi:2011uz,Donagi:2011va,Closset:2015ohf}, 
do not contribute
to quantum sheaf cohomology or A/2-model correlation functions, so we
omit nonlinear deformations.)
The (2,2) locus is given by the special case
\begin{displaymath}
A = I, \: \: \:
B = 0, \: \: \:
\gamma_1 = n, \: \: \:
\gamma_2 = 1, 
\alpha_1 = 0, \: \: \:
\alpha_2 = 1.
\end{displaymath}

We have the same constraints on fields from D terms as on the (2,2) locus,
namely
\begin{displaymath}
Y_1 + Y_2 + n Y_s = t_1, \: \: \:
Y_s + Y_t = t_2,
\end{displaymath}
where $Y_{1,2}$ are dual to the $x$'s, $Y_3$ is dual to $s$,
and $Y_4$ is dual to $t$.
We can solve them by taking
\begin{displaymath}
\tilde{t}_1 = 0, \: \: \:
\tilde{t}_2 = t_1, \: \: \:
\tilde{t}_s = 0, \: \: \:
\tilde{t}_t = t_2,
\end{displaymath}
\begin{displaymath}
(V^A_i) \: = \: \left[ \begin{array}{cccc}
1 & -1 & 0 & 0 \\
0 & -n & 1 & -1
\end{array} \right],
\end{displaymath}
so that
\begin{displaymath}
Y_1 = \theta, \: \: \:
Y_2 = t_1 - \theta - n \tilde{\theta}, \: \: \:
G_1 = F_1 = - F_2 - n G_2,
\end{displaymath}
\begin{displaymath}
Y_3 = \tilde{\theta}, \: \: \:
Y_4 = t_2 - \tilde{\theta}, \: \: \:
G_2 = F_3 = - F_4.
\end{displaymath}

\subsubsection{First choice of $S$}

We take the submatrix $S \subset Q$ to correspond to the
first and third columns of the
charge matrix $Q$, {\it i.e.}
\begin{displaymath}
S \: = \: \left[ \begin{array}{cc}
1 & n \\ 0 & 1 \end{array} \right].
\end{displaymath}

The allowed deformations are
\begin{displaymath}
(A_{ij}) \: = \: \left[ \begin{array}{cccc}
1 & 0 & 0 & 0 \\
A_{21} & A_{22} & A_{23} & A_{24} \\
0 & 0 & 1 & 0 \\
A_{41} & A_{42} & A_{43} & A_{44} 
\end{array} \right].
\end{displaymath}
To find the corresponding elements of $\tilde{A}$, $\tilde{B}$,
$\gamma_{1,2}$, $\alpha_{1,2}$, we compare the $E$'s.
For the deformations defined by $A_{ij}$,
\begin{eqnarray*}
E_1 & = & \sum_a Q_a^a \sigma_a \phi_1 \: = \: \sigma_1 \phi_1, \\
E_2 & = & \sum_{j, a} A_{2j} Q_j^a \sigma_a \phi_2 ,\\
& = & \left( A_{21} \sigma_1 + A_{22} \sigma_1 + A_{23}(n \sigma_1 + \sigma_2)
+ A_{24} \sigma_2 \right) \phi_2, \\
E_s & = & (n \sigma_1 + \sigma_2) s, \\
E_t & = & \left( A_{41} \sigma_1 + A_{42} \sigma_1 + A_{43}(n \sigma_1 +
\sigma_2) + A_{44} \sigma_2 \right) t,
\end{eqnarray*}
whereas for the bundle deformation parameters,
\begin{eqnarray*}
E_1 & = & \left( \tilde{A}_{11} \phi_1 + \tilde{A}_{12} \phi_2 \right)
\sigma_1 + \left( \tilde{B}_{11} \phi_1 + \tilde{B}_{12} \phi_2 \right)
\sigma_2, \\
E_2 & = & \left( \tilde{A}_{21} \phi_1 + \tilde{A}_{22} \phi_2 \right) \sigma_1
+ \left( \tilde{B}_{21} \phi_1 + \tilde{B}_{22} \phi_2 \right) \sigma_2, \\
E_s & = & \gamma_1 s \sigma_1 + \gamma_2 s \sigma_2, \\
E_t & = & \alpha_1 t \sigma_1 + \alpha_2 t \sigma_2, 
\end{eqnarray*}
from which we read off
\begin{displaymath}
\tilde{A} = \left[ \begin{array}{cc}
1 & 0 \\
0 & A_{21} + A_{22} + n A_{23} \end{array} \right], \: \: \:
\tilde{B} = \left[ \begin{array}{cc}
0 & 0 \\ 0 & A_{23} + A_{24} \end{array} \right],
\end{displaymath}
\begin{displaymath}
a = \det \tilde{A} = A_{21}+A_{22}+nA_{23}, \: \: \:
b = \det \tilde{B} = 0, \: \: \:
\mu = A_{23}+A_{24},
\end{displaymath}
\begin{displaymath}
\gamma_1 = n, \: \: \:
\gamma_2 = 1, \: \: \:
\alpha_1 = A_{41} + A_{42} + n A_{43}, \: \: \:
\alpha_2 = A_{43} + A_{44}.
\end{displaymath}

Next, let us construct the mirror.  From formula~(\ref{eq:d-defn}),
\begin{displaymath}
(D^A_{i_S}) \: = \: \left[ \begin{array}{cc}
A_{21}+A_{22}-nA_{24} -1 & A_{23}+A_{24} \\
n(A_{21}+A_{22}-nA_{24}) + (A_{41}+A_{42}-nA_{44}) &
n(A_{23}+A_{24}) + (A_{43}+A_{44})-1
\end{array} \right],
\end{displaymath}

From equation~(\ref{eq:02mirror-w}), the proposed mirror superpotential is then
\begin{eqnarray*}
W & = & - G_1 \left( e^{-Y_1} - e^{-Y_2} + (A_{21}+A_{22}-nA_{24}-1)
e^{-Y_1} + (A_{23}+A_{24}) e^{-Y_3} \right) \\
& & - G_2 \left( -n e^{-Y_2} + e^{-Y_3} - e^{-Y_4} +
\left( n(A_{21}+A_{22}-nA_{24}) + (A_{41}+A_{42}-nA_{44}) \right) e^{-Y_1}
\right. \\
& & \hspace*{0.6in} \left.
 + \left( n(A_{23}+A_{24}) + (A_{43}+A_{44})-1 \right) e^{-Y_3}
\right),
\\
& = & - G_1\left( (A_{21}+A_{22}-nA_{24})X_1 - \frac{q_1}{X_1 X_3^n} 
+ (A_{23}+A_{24}) X_3 \right) \\
& & - G_2 \left( -n \frac{q_1}{X_1 X_3^n} + 
\left(n(A_{23}+A_{24}) + (A_{43}+A_{44})\right)X_3
 - \frac{q_2}{X_3}
\right. \\
& & \hspace*{0.6in} 
 + \left(
 n(A_{21}+A_{22}-nA_{24}) + (A_{41}+A_{42}-nA_{44})\right) X_1
\Bigg),
\end{eqnarray*}
where $X_i = \exp(-Y_i)$,
with operator mirror map~(\ref{eq:02mirror})
\begin{eqnarray*}
X_1 & \leftrightarrow & \sigma_1, \\
X_2 = \frac{q_1}{X_1 X_3^n}& \leftrightarrow & (A_{21}+A_{22})\sigma_1 + A_{23}(n \sigma_1 +
\sigma_2) + A_{24} \sigma_2 , \\
X_3 & \leftrightarrow & n \sigma_1 + \sigma_2 , \\
X_4 = \frac{q_2}{X_3}
& \leftrightarrow & (A_{41}+A_{42}) \sigma_1 + A_{43}(n \sigma_1 +
\sigma_2) + A_{44} \sigma_2.
\end{eqnarray*}
Note that the operator mirror map relations for $X_2$, $X_4$ are
consequences of the equations of motion $\partial W/\partial G_A = 0$.

For these deformations, the quantum sheaf cohomology ring is given by
\cite{McOrist:2008ji,Donagi:2011uz,Donagi:2011va,Closset:2015ohf}
\begin{displaymath}
Q_{(k)} Q_{(s)}^n \: = \: q_1, \: \: \:
Q_{(s)} Q_{(t)} \: = \: q_2,
\end{displaymath}
where
\begin{displaymath}
Q_{(k)} \: = \: (A_{21} + A_{22} + n A_{23} ) \sigma_1^2 \: + \:
(A_{23}+A_{24}) \sigma_1 \sigma_2, 
\end{displaymath}
\begin{displaymath}
Q_{(s)} \: = \: n \sigma_1 + \sigma_2, \: \: \:
Q_{(t)} \: = \: (A_{41}+A_{42}+nA_{43}) \sigma_1 +
(A_{43}+A_{44}) \sigma_2.
\end{displaymath}
It is straightforward to check that these relations are 
implied by the mirror map equations above.

A proposal was made in \cite{Chen:2017mxp} for the Toda dual to a (GLSM for
a) Hirzebruch surface.  Briefly, the mirror superpotential had the form
\begin{displaymath}
W \: = \: - G_1 J_1 - G_2 J_2
\end{displaymath}
for \cite{Chen:2017mxp}[section 4.2]  
\begin{align}
J_1 &=  a X_1 + \mu_{AB} (X_3 - n X_1) + b \frac{(X_3 - n X_1)^2}{X_1}  \nonumber \\
& \qquad \qquad \qquad \qquad \qquad \qquad - q_1 X_1^{-1} \left( \gamma_1 X_1 + \gamma_2 (X_3 - n X_1) \right)^{-n}, \label{eq:hir-02-1}\\
J_2 &= n \left(  a X_1 + \mu_{AB} (X_3 - n X_1) 
+ b \frac{(X_3 - n X_1)^2}{X_1} \right) \nonumber  \\
& \qquad \qquad \qquad  - \frac{ n q_1 }{ X_1 
\left( \gamma_1 X_1 + \gamma_2 (X_3 - n X_1) \right)^{n}  }
- \frac{q_2}{X_3} \nonumber \\
&\qquad \qquad \qquad \qquad + \frac{
\left( \gamma_1 X_1 + \gamma_2 (X_3 - n X_1) \right) \left( \alpha_1 X_1 + \alpha_2 (X_3 - n X_1) \right) }{X_3}, \label{eq:hir-02-2}
\end{align}
with operator mirror map 
\begin{displaymath}
X_1 \: \leftrightarrow \: \sigma_1, \: \: \:
X_3 \: \leftrightarrow \: n \sigma_1 + \sigma_2.
\end{displaymath}

It is straightforward to check that the proposal of \cite{Chen:2017mxp},
reviewed above,
specializes to our proposal here.

\subsubsection{Second choice of $S$}

Next, consider the case that the submatrix $S \subset Q$ is taken to
correspond to the first and fourth columns of $Q$, {\it i.e.}
\begin{displaymath}
S \: = \: \left[ \begin{array}{cc}
1 & 0 \\ 0 & 1 \end{array} \right].
\end{displaymath}
The allowed deformations are
\begin{displaymath}
(A_{ij}) \: = \: \left[ \begin{array}{cccc}
1 & 0 & 0 & 0 \\
A_{21} & A_{22} & A_{23} & A_{24} \\
A_{31} & A_{32} & A_{33} & A_{34} \\
0 & 0 & 0 & 1 
\end{array} \right]
\end{displaymath}
Proceeding as before, the corresponding $\tilde{A}$, $\tilde{B}$,
$\gamma_{1,2}$, $\alpha_{1,2}$ are given by
\begin{displaymath}
\tilde{A} \: = \: \left[ \begin{array}{cc}
1 & 0 \\
0 & A_{21}+A_{22}+n A_{23} \end{array} \right], \: \: \:
\tilde{B} \: = \: \left[ \begin{array}{cc}
0 & 0 \\
0 & A_{23} + A_{24} \end{array} \right], 
\end{displaymath}
\begin{displaymath}
a = \det \tilde{A} = A_{21}+A_{22}+ n A_{23}, \: \: \:
b = \det \tilde{B} = 0, \: \: \:
\mu = A_{23} + A_{24},
\end{displaymath}
\begin{displaymath}
\gamma_1 = A_{31}+A_{32}+ n A_{33}, \: \: \:
\gamma_2 = A_{33} + A_{34},
\end{displaymath}
\begin{displaymath}
\alpha_1 = 0, \: \: \:
\alpha_2 = 1.
\end{displaymath}

Next, let us construct the mirror.  From formula~(\ref{eq:d-defn}),
\begin{displaymath}
(D^A_{i_S}) \: = \: \left[ \begin{array}{cc}
A_{21}+A_{22}+ n A_{23} - 1 & A_{23}+A_{24} \\
n (A_{21}+A_{22} + n A_{23} ) - (A_{31}+A_{32}+nA_{33}) &
n(A_{23}+A_{24}) - (A_{33}+A_{34}-1)
\end{array} \right].
\end{displaymath}

From equation~(\ref{eq:02mirror-w}), the proposed mirror superpotential is then
\begin{eqnarray*}
W & = & - G_1\left( e^{-Y_1} - e^{-Y_2} + (A_{21}+A_{22}+ n A_{23} - 1)
e^{-Y_1} + (A_{23}+A_{24}) e^{-Y_4}
\right) \\
& & - G_2\left( -n e^{-Y_2} + e^{-Y_3} - e^{-Y_4} +
\left( n (A_{21}+A_{22} + n A_{23} ) - (A_{31}+A_{32}+nA_{33})
\right) e^{-Y_1}
\right. \\
& & \hspace*{0.6in} \left.
 +
\left( n(A_{23}+A_{24}) - (A_{33}+A_{34}-1) \right) e^{-Y_4}
\right), \\
& = & - G_1 \left(
( A_{21}+A_{22} + n A_{23}) X_1 -  \frac{q_1}{q_2^n} \frac{X_4^n}{X_1} +
(A_{23}+A_{24}) X_4 \right)
\\
& & - G_2 \left( -n \frac{q_1}{q_2^n} \frac{X_4^n}{X_1}
+ \frac{q_2}{X_4}  
+ \left( n(A_{23}+A_{24}) - (A_{33}+A_{34}) \right) X_4
\right. \\
& & \hspace*{0.5in}
+ \left( n (A_{21}+A_{22} + n A_{23} ) - (A_{31}+A_{32}+nA_{33})
\right) X_1 
\Bigg),
\end{eqnarray*}
where $X_i = \exp(-Y_i)$,
with operator mirror map~(\ref{eq:02mirror})
\begin{eqnarray*}
X_1 & \leftrightarrow & \sigma_1, \\
X_2 = \frac{q_1}{q_2^n} \frac{X_4^n}{X_1}
& \leftrightarrow & (A_{21}+A_{22}) \sigma_1 + A_{23}(n \sigma_1 +
\sigma_2) + A_{24} \sigma_2, \\
X_3  = \frac{q_2}{X_4}
& \leftrightarrow & (A_{31}+A_{32}) \sigma_1 + A_{33}(n \sigma_1 +
\sigma_2) + A_{34} \sigma_2, \\
X_4 & \leftrightarrow & \sigma_2.
\end{eqnarray*}
The operator mirror map relation for $X_2$ is a consequence of $\partial W/
\partial G_1=0$, and the operator mirror map relation for $X_3$ is a consequence
of that plus $\partial W/\partial G_2=0$.

A second proposal was made in \cite{Chen:2017mxp} for the Toda dual to a
(GLSM for a) Hirzebruch surface, in which the mirror superpotential had the
form
\begin{displaymath}
W \: = \: - G_1 J_1 - G_2 J_2,
\end{displaymath}
for \cite{Chen:2017mxp}[section 4.2]
\begin{align}
J_1 &=  \left( a X_1 + \mu_{AB} X_4 + b \frac{X_4^2}{X_1} \right) 
- \frac{q_1}{q_2^n} \frac{(\alpha_1 X_1 + \alpha_2 X_4)^n}{X_1}, \label{eq:hir-02-3} \\
J_2 &=  
-n  \left( a X_1 + \mu_{AB} X_4 + b \frac{X_4^2}{X_1}  
- \frac{q_1}{q_2^n} \frac{(\alpha_1 X_1 + \alpha_2 X_4)^n}{X_1}
\right)   \nonumber
\\
& \hspace*{0.2in}
+  \left( \alpha_2 \gamma_2 X_4 + \gamma_1 \alpha_1 \frac{X_1^2}{X_4} 
+ (\gamma_1 \alpha_2 + \gamma_2 \alpha_1) X_1
\right) 
- \frac{q_2}{X_4}. \label{eq:hir-02-4}
\end{align}
with operator mirror map
\begin{displaymath}
X_1 \: \leftrightarrow \: \sigma_1, \: \: \:
X_4 \: \leftrightarrow \: \sigma_2.
\end{displaymath}
It is straightforward to check that this proposal of
\cite{Chen:2017mxp} specializes to our proposal.

\subsection{$dP_2$}

Let us now consider the del Pezzo surface $dP_2$, corresponding to
${\mathbb P}^2$ blown up at two points, which was also considered
in \cite{Chen:2017mxp}.
The charge matrix of the GLSM for the chiral fields is of the form
\begin{displaymath}
\left[ \begin{array}{ccccc}
1 & 1 & 1 & 0 & 0\\
0 & 0 & 1 & 1 & 0 \\
1 & 0 & 0 & 0 & 1
\end{array} \right],
\end{displaymath}
and deformations of the tangent bundle are described mathematically as the
cokernel ${\cal E}$ of the short exact sequence
\begin{equation*}
\begin{split}
0 \: \longrightarrow {\cal O}^3 \: \xrightarrow{*} \: {\cal O}(1,0,1) \oplus {\cal O}(1,0,0) \oplus {\cal O}(1,1,0) \oplus  {\cal O}(0,1,0) \oplus {\cal O}(0,0,1) \: \\
\quad  \longrightarrow \: {\cal E} \: \longrightarrow 0 ,
\end{split}
\end{equation*}
where
\begin{displaymath}
* \: = \: \left[ \begin{array}{ccc}
\alpha_1 \phi_1 & \alpha_2  \phi_2 & \alpha_3 \phi_3 \\
\beta_1 \phi_1 & \beta_2  \phi_2 & \beta_3 \phi_3 \\
\gamma_1 \phi_1 & \gamma_2  \phi_2 & \gamma_3 \phi_3 \\
\delta_1 \phi_1 & \delta_2  \phi_2 & \delta_3 \phi_3 \\
\epsilon_1 \phi_1 & \epsilon_2  \phi_2 & \epsilon_3 \phi_3 
\end{array} \right].
\end{displaymath}
The (2,2) locus is given by the special case
\begin{align*}
 &\alpha_1 = 1, \quad \alpha_2 = 0, \quad \alpha_3 = 1, \\
 &\beta_1 =1, \quad \beta_2 = \beta_3 = 0, \\
 &\gamma_1 = \gamma_2 =1, \quad \gamma_3 = 0, \\
 &\delta_1 = 0, \quad \delta_2 = 1, \quad \delta_3 = 0, \\
 &\epsilon_1 = \epsilon_2 = 0, \quad \epsilon_3 = 1.
\end{align*}

We will take
\begin{displaymath}
(V^A_i) \: = \: \left[ \begin{array}{ccccc}
1 & -1 & 0 & 0 & -1 \\
0 & -1 & 1 & -1 & 0
\end{array} \right].
\end{displaymath}

\subsubsection{First choice of $S$}

For our first choice of $S$, we will take the first, third, and fifth
columns of the charge matrix, so that
\begin{displaymath}
S \: = \: \left[ \begin{array}{ccc}
1 & 1 & 0 \\
0 & 1 & 0 \\
1 & 0 & 1
\end{array} \right].
\end{displaymath}
With this choice of $S$, the allowed deformations are
\begin{displaymath}
(A_{ij}) \: = \: \left[ \begin{array}{ccccc}
1 & 0 & 0 & 0 & 0 \\
A_{21} & A_{22} & A_{23} & A_{24} & A_{25} \\
0 & 0 & 1 & 0 & 0 \\
A_{41} & A_{42} & A_{43} & A_{44} & A_{45} \\
0 & 0 & 0 & 0 & 1
\end{array}  \right].
\end{displaymath}
We compute the corresponding bundle deformation parameters by comparing $E$'s:
\begin{eqnarray*}
E_1 & = & ( \alpha_1 \sigma_1 + \alpha_2 \sigma_2 + \alpha_3 \sigma_3)
\phi_1 
\: = \: (\sigma_1 + \sigma_3) \phi_1, \\
E_2 & = &  ( \beta_1 \sigma_1 + \beta_2 \sigma_2 + \beta_3 \sigma_3)
\phi_2 , \\
& = & 
\left( A_{21}(\sigma_1 + \sigma_3) + A_{22} \sigma_1 + A_{23}(\sigma_1 +
 \sigma_2) + A_{24} \sigma_2 + A_{25} \sigma_3 \right) \phi_2, \\
E_3 & = &  ( \gamma_1 \sigma_1 + \gamma_2 \sigma_2 + \gamma_3 \sigma_3)
\phi_3 
\: = \: (\sigma_1 + \sigma_2) \phi_3, \\
E_4 & = & ( \delta_1 \sigma_1 + \delta_2 \sigma_2 + \delta_3 \sigma_3)
\phi_4, \\
& = & \left( A_{41}(\sigma_1 + \sigma_3) + A_{42} \sigma_1 + A_{43}(
\sigma_1 + \sigma_2) + A_{44} \sigma_2 + A_{45} \sigma_3 \right) \phi_4, \\
E_5 & = &  ( \epsilon_1 \sigma_1 + \epsilon_2 \sigma_2 + \epsilon_3 \sigma_3)
\phi_5 \: = \:
\sigma_3 \phi_5,
\end{eqnarray*}
from which we find
\begin{displaymath}
\vec{\alpha} = (1,0,1), \: \: \:
\vec{\gamma} = (1,1,0), \: \: \:
\vec{\epsilon} = (0,0,1),
\end{displaymath}
\begin{displaymath}
\vec{\beta} = (A_{21}+A_{22}+A_{23}, A_{23}+A_{24}, A_{21}+A_{25}), 
\end{displaymath}
\begin{displaymath}
\vec{\delta} = (A_{41}+A_{42}+A_{43}, A_{43}+A_{44}, A_{41}+A_{45}).
\end{displaymath}

From formula~(\ref{eq:d-defn}), we have
\begin{displaymath}
(D^A_{i_S}) \: = \: \left[ \begin{array}{ccc}
A_{21}+A_{22}-1-A_{24} & A_{23}+A_{24} & -A_{22}+A_{24}+A_{25}+1 \\
{\begin{array}{l} A_{21}+A_{22}-A_{24}+\\  \qquad A_{41}+A_{42}-A_{44}
\end{array} } &
A_{23}+A_{24}+A_{43}+A_{44}-1 &
{\begin{array}{l} -A_{22} + A_{24} + A_{25}\\  \qquad 
- A_{42}+A_{44}+A_{45}
\end{array} }
\end{array} \right].
\end{displaymath}
The proposed mirror superpotential~(\ref{eq:02mirror-w}) is
then
\begin{eqnarray*}
W & = & - G_1 \left[ (A_{21}+A_{22}-A_{24}) X_1 - \frac{q_1}{X_1 X_3}
+ (A_{23}+A_{24})X_3 + (-A_{22} + A_{24} + A_{25}) X_5 \right] \\
& & - G_2 \left[ (A_{21}+A_{22}-A_{24}+A_{41}+A_{42}-A_{44}) X_1
- \frac{q_1}{X_1 X_3} +
(A_{23}+A_{24}+A_{43}+A_{44}) X_3
\right. \\
& & \hspace*{0.6in} \left.
 - \frac{q_2}{X_3} +
(-A_{22}+A_{24}+A_{25}-A_{42}+A_{44}+A_{45}) X_5 \right],
\end{eqnarray*}
with operator mirror map~(\ref{eq:02mirror})
\begin{eqnarray*}
X_1 & \leftrightarrow & \sigma_1 + \sigma_3, \\
X_2  = \frac{q_1}{X_1 X_3} & \leftrightarrow & (A_{21}+A_{22}+A_{23}) \sigma_1
+ (A_{23}+A_{24}) \sigma_2 + (A_{21}+A_{25}) \sigma_3, \\
X_3 & \leftrightarrow & \sigma_1 + \sigma_2, \\
X_4 = \frac{q_2}{X_3} & \leftrightarrow & (A_{41}+A_{42}+A_{43}) \sigma_1 +
(A_{43}+A_{44}) \sigma_2 + (A_{41}+A_{45})\sigma_3, \\
X_5 = \frac{q_3}{X_1}& \leftrightarrow & \sigma_3.
\end{eqnarray*}
(It is straightforward to check that the nontrivial operator mirror map
relations above are equivalent to the equations of motion derived
from $\partial W/\partial G_A = 0$.)

Now, let us compare to the first (0,2) mirror proposal for $dP_2$ in
\cite{Chen:2017mxp}[section 3.2.2].  In their notation
\begin{displaymath}
\alpha \cdot X = X_1, \: \: \:
\gamma \cdot X = X_3, \: \: \:
\epsilon \cdot X = X_5,
\end{displaymath}
\begin{eqnarray*}
\beta \cdot X & = & (A_{21} + A_{22}-A_{24}) X_1 + (A_{23}+A_{24}) X_3
+ (A_{24}+A_{25}-A_{22}) X_5, \\
\delta \cdot X & = & (A_{41} + A_{42} - A_{44}) X_1 +
(A_{43}+A_{44}) X_3 +
(A_{44}+A_{45}-A_{22}) X_5,
\end{eqnarray*}
and
\begin{eqnarray*}
J_1 & = & - \frac{q_1}{X_1 X_3} + Z \frac{q_3}{X_1 X_5} + X_5
+ \beta \cdot X, \\
J_3 & = & - \frac{q_2}{X_3} - \frac{q_1}{X_1 X_3} + \beta \cdot X +
\delta \cdot X, \\
J_5 & = & X_5 + Z \frac{q_3}{X_1 X_5}, \\
J_Z & = & 1 - \frac{q_3}{X_1 X_5}.
\end{eqnarray*}
Solving $J_Z=0$, we get $X_5 = q_3/X_1$,
and solving $J_5=0$, we get $Z = - X_5$.
Plugging in, we are left with two independent $J$'s:
\begin{eqnarray*}
J_1 & = & - \frac{q_1}{X_1 X_3} + (A_{21}+A_{22}-A_{24}) X_1 + (A_{23}+A_{24})
X_3 \\
& &\hspace*{0.2in}
 + (A_{24}+A_{25}-A_{22}) \frac{q_3}{X_1}, \\
J_3 & = & -\frac{q_2}{X_3} - \frac{q_1}{X_1 X_3} +
(A_{21}+A_{22} - A_{24} + A_{41} + A_{42} - A_{44}) X_1 \\
& & + (A_{23}+A_{24} + A_{43}+A_{44}) X_3 +
(A_{24}+A_{25}-A_{22} + A_{44}+A_{45}-A_{42}) 
\frac{q_3}{X_1},
\end{eqnarray*}
which precisely matches the prediction of our proposal.

Finally, let us compare the quantum sheaf cohomology ring relations.
There are three quantum sheaf cohomology ring relations, but only two
relations appearing above in the operator mirror map and equations of motion.
Specifically, in this case, the quantum sheaf cohomology ring relations
\cite{McOrist:2008ji,Donagi:2011uz,Donagi:2011va} take the form
\begin{eqnarray}
(\sigma_1 + \sigma_3) (\sigma_1 + \sigma_2) Q_{(2)} & = & q_1, \\
(\sigma_1 + \sigma_2) Q_{(4)} & = & q_2, \\
(\sigma_1 + \sigma_3) \sigma_3 & = & q_3,
\end{eqnarray}
where
\begin{eqnarray*}
Q_{(2)} & = & (A_{21} + A_{22} + A_{23}) \sigma_1 + (A_{23}+A_{24})\sigma_2
+ (A_{21}+A_{25}) \sigma_3, \\
Q_{(4)} & = & (A_{41}+A_{42}+A_{43}) \sigma_1 + (A_{43}+A_{44}) \sigma_2 
+ (A_{41}+A_{45}) \sigma_3.
\end{eqnarray*}
The first two quantum sheaf cohomology ring relations correspond to two of
the operator mirror map statements.  The third is realized on the mirror as the
relation
\begin{displaymath}
X_1 X_5 \: = \: q_3,
\end{displaymath}
which is a consequence of the D terms, and so automatic.

\subsubsection{Second choice of $S$}

For our second choice of $S$, we will take the second, fourth, and fifth
columns of the charge matrix, so that $S$ is the identity.  With that choice
of $S$, the allowed deformations are
\begin{displaymath}
(A_{ij}) \: = \: \left[ \begin{array}{ccccc}
A_{11} & A_{12} & A_{13} & A_{14} & A_{15} \\
0 & 1 & 0 & 0 & 0 \\
A_{31} & A_{32} & A_{33} & A_{34} & A_{35} \\
0 & 0 & 0 & 1 & 0 \\
0 & 0 & 0 & 0 & 1
\end{array} \right].
\end{displaymath}
To find the corresponding bundle deformation parameters, we compare
the $E$'s:
\begin{eqnarray*}
E_1 & = & ( \alpha_1 \sigma_1 + \alpha_2 \sigma_2 + \alpha_3 \sigma_3)
\phi_1 , \\
& = &
\left( A_{11} (\sigma_1 + \sigma_3) + A_{12} \sigma_1 +
A_{13} (\sigma_1 + \sigma_2) + A_{14} \sigma_2 +
A_{15} \sigma_3 \right) \phi_1, \\
E_2 & = &  ( \beta_1 \sigma_1 + \beta_2 \sigma_2 + \beta_3 \sigma_3)
\phi_2 \: = \:
\sigma_1 \phi_2, \\
E_3 & = &  ( \gamma_1 \sigma_1 + \gamma_2 \sigma_2 + \gamma_3 \sigma_3)
\phi_3 , \\
& = &
\left( A_{31} (\sigma_1 + \sigma_3) + A_{32} \sigma_1 +
A_{33} (\sigma_1 + \sigma_2) + A_{34} \sigma_2 +
A_{35} \sigma_3 \right) \phi_3, \\
E_4 & = & ( \delta_1 \sigma_1 + \delta_2 \sigma_2 + \delta_3 \sigma_3)
\phi_4 \: = \:
\sigma_2 \phi_4, \\
E_5 & = &  ( \epsilon_1 \sigma_1 + \epsilon_2 \sigma_2 + \epsilon_3 \sigma_3)
\phi_5 \: = \:
\sigma_3 \phi_5,
\end{eqnarray*}
from which we conclude
\begin{displaymath}
\vec{\alpha} \: = \: ( A_{11}+A_{12}+A_{13}, A_{13}+A_{14}, A_{11}+A_{15}),
\end{displaymath}
\begin{displaymath}
\vec{\beta} = (1,0,0), \: \: \:
\vec{\delta} = (0,1,0), \: \: \:
\vec{\epsilon} = (0,0,1),
\end{displaymath}
\begin{displaymath}
\vec{\gamma} \: = \: ( A_{31}+A_{32}+A_{33}, A_{33}+A_{34}, A_{31}+A_{35}) .
\end{displaymath}

Next, we construct the mirror.  From formula~(\ref{eq:d-defn}),
we have
\begin{displaymath}
(D^A_{i_S}) \: = \: - \left[ \begin{array}{ccc}
A_{11}+A_{12}+A_{13}-1 & A_{13}+A_{14} & A_{11}+A_{15}-1 \\
A_{31}+A_{32}+A_{33}-1 & A_{33}+A_{34}-1 & A_{31}+A_{35}
\end{array} \right],
\end{displaymath}
then the proposed mirror superpotential~(\ref{eq:02mirror-w}) is
\begin{eqnarray*}
W & = & - G_1 \left( \frac{q_1}{q_2}\frac{X_4}{X_2}
 - (A_{11}+A_{12}+A_{13}) X_2 - (A_{13}+A_{14}) X_4
- (A_{11}+A_{15}) X_5 \right) \\
& & - G_2 \left( - (A_{31}+A_{32}+A_{33}) X_2 + \frac{q_2}{X_4}
 - (A_{33}+A_{34})X_4
- (A_{31}+A_{35}) X_5 \right),
\end{eqnarray*}
where $X_i = \exp(-Y_i)$.

The operator mirror map~(\ref{eq:02mirror}) is then given by
\begin{eqnarray*}
X_1 = \frac{q_1}{q_2}\frac{X_4}{X_2} & \leftrightarrow & A_{11}(\sigma_1+\sigma_3) + A_{12} \sigma_1 +
A_{13}(\sigma_1+\sigma_2) + A_{14} \sigma_2 + A_{15} \sigma_3, \\
X_2 & \leftrightarrow & \sigma_1 , \\
X_3 = \frac{q_2}{X_4} & \leftrightarrow & A_{31}(\sigma_1 + \sigma_3) + A_{32} \sigma_1 +
A_{33}(\sigma_1 + \sigma_2) + A_{34} \sigma_2 + A_{35} \sigma_3, \\
X_4 & \leftrightarrow & \sigma_2, \\
X_5 & \leftrightarrow & \sigma_3.
\end{eqnarray*}

From \cite{McOrist:2008ji,Donagi:2011uz,Donagi:2011va}, 
the quantum sheaf cohomology relations in this model are
given by
\begin{eqnarray}
\sigma_1 Q_{(1)} Q_{(3)} & = & q_1,  \label{eq:dp2:ex1:qsc1}\\
\sigma_2 Q_{(3)} & = & q_2, \label{eq:dp2:ex1:qsc2} \\
\sigma_3 Q_{(1)} & = & q_3,  \label{eq:dp2:ex1:qsc3}
\end{eqnarray}
where
\begin{eqnarray*}
Q_{(1)} & = & (A_{11}+A_{12}+A_{13}) \sigma_1 
+ (A_{13}+A_{14}) \sigma_2 +
(A_{11}+A_{15}) \sigma_3, \\
Q_{(3)} & = & (A_{31}+A_{32}+A_{33}) \sigma_1 +
(A_{33}+A_{34}) \sigma_2 + (A_{31}+A_{35}) \sigma_3.
\end{eqnarray*}
This first two quantum sheaf cohomology
relations can be seen to correspond to both the operator mirror map
relations and the equations of motion.  The third relation 
(\ref{eq:dp2:ex1:qsc3}) is effectively redundant on the mirror.  To see
this, note that it corresponds to the statement
\begin{displaymath}
X_5 Q_{(1)} \: = \: q_3.
\end{displaymath}
However, $X_5 = q_3/X_1$ from the D-term relations, so this is equivalent to
\begin{displaymath}
X_1 \: = \: Q_{(1)},
\end{displaymath}
which is the first mirror-map relation.

\section{Hypersurfaces in toric varieties}
\label{sect:hyp}

\subsection{General aspects}

An extension of ordinary mirror symmetry from toric varieties to hypersurfaces
therein was discussed in \cite{Hori:2000kt} and further justified in 
\cite{Aganagic:2004yn}.
Let us begin our discussion here by very briefly reviewing this for the
special case of the (2,2) quintic in ${\mathbb P}^4$.  One begins with the
Toda dual of Tot (${\cal O}(-5) \rightarrow {\mathbb P}^4$), the ambient
GLSM for the quintic albeit with vanishing superpotential.  The Toda dual
is defined by a (twisted) superpotential of the form
\begin{displaymath}
W \: = \: \sum_{i=1}^5 \exp(-Y_i) \: + \: \exp(-Y_p),
\end{displaymath}
where the D term constraint requires
\begin{displaymath}
- 5 Y_p \: + \: \sum_{i=1}^n Y_i \: = \: t.
\end{displaymath}
The effect of dualizing the GLSM with a nonzero superpotential,
according to \cite{Hori:2000kt,Aganagic:2004yn} is to change the
fundamental fields from $Y_i$ to $X_i \equiv \exp(-Y_i/5)$, 
which also introduces
${\mathbb Z}_5$ orbifolds.  As a result, after eliminating $Y_p$
with the D term constraint and the change of variables,
the mirror superpotential becomes (${\mathbb Z}_5$ orbifolds of)
\begin{displaymath}
W \: = \: \sum_i X_i^5 \: + \: q \prod_i X_i.
\end{displaymath}
The fact that the B-twisted mirror sits at a Landau-Ginzburg point
reflects the fact that the B model is independent of K\"ahler moduli,
and so topological field theory computations can be computed at any
point on the K\"ahler moduli space -- the Landau-Ginzburg orbifold
point in the moduli space of a hypersurface being a 
convenient example.

Before considering (0,2) analogues, let us quickly outline a 
formal justification for the change of variables above, in TFT
correlation functions.
We claim that at the level of untwisted-sector correlation functions,
it is formally equivalent to insertions needed to
restrict to the hypersurface.  The argument below omits questions of
counting and degeneracy of vacua, as well as how the kinetic terms of
the fields change, but is sufficiently tantalizing that we mention it here.

Begin with a (2,2) supersymmetric GLSM for 
Tot (${\cal O}(-5) \rightarrow {\mathbb P}^4$),
labelled the $V^+$ model in \cite{Morrison:1994fr}.
To compute a correlation function matching one on the hypersurface,
in principle one should insert $(-)(-5 \sigma)^2$ 
(see {\it e.g.} \cite{Morrison:1994fr}[equ'n (5.8)]).
Correlation functions in the untwisted sector of 
the mirror B-twisted Landau-Ginzburg 
model, corresponding to certain computations on the hypersurface, should
then have the form
\begin{displaymath}
\langle {\cal O}_1 \cdots {\cal O}_n \rangle_{\rm quintic} \: = \:
\langle {\cal O}_1 \cdots {\cal O}_n (-)(- 5 \sigma)^2 \rangle_{V^+} \: \propto \:
\sum_{\rm vacua} \frac{ {\cal O}_1 \cdots {\cal O}_n \exp(-2Y_p) }{
\det(\partial_{Y_i} \partial_{Y_j} W ) }.
\end{displaymath}
where we have used the mirror map
\begin{displaymath}
\exp(-Y_p) = e^{-t/5} \prod_i \exp(-Y_i/5) \: \leftrightarrow \: -5 \sigma.
\end{displaymath}
Now, we change variables from $Y_i$ to $X_i \equiv \exp(-Y_i/5)$.
Formally, in the denominator, 
\begin{eqnarray*}
\frac{\partial}{\partial Y_i} W & = & \frac{\partial X_j}{\partial Y_i}
\frac{\partial W}{\partial X_j} \: = \: - \frac{1}{5} X_i 
\frac{\partial W}{\partial X_i}, \\
\frac{\partial^2 W}{\partial Y_i \partial Y_j} & = & \left( - \frac{1}{5}
\right)^2 X_i X_j \frac{\partial^2 W}{\partial X_i \partial X_j}
\mbox{  (no sum on $i$, $j$)},
\end{eqnarray*}
along the critical locus.  Now, strictly speaking, after the change
of variables, the critical locus becomes degenerate, the second
derivative of the superpotential vanishes at the critical locus.
One could attempt to solve this by turning on twisted masses,
possible for certain special superpotentials, but as our aim here is
a formal observation, we shall move on.  The point we wish to make
is that formally, if one plugs into the expression for
correlation functions, then glossing over questions of number and
degeneracy of vacua, 
\begin{displaymath}
\langle {\cal O}_1 \cdots {\cal O}_n \rangle_{\rm quintic} \: = \:
\sum_{\rm vacua} \frac{ {\cal O}_1 \cdots {\cal O}_n \exp(-2Y_p) }{
(-1/5)^{10} \left( \prod_i X_i\right)^2 \det\left(\partial_{X_i} 
\partial_{X_j} W \right) 
},
\end{displaymath}
then along the vacua, the factor of $\exp(-2Y_p) = e^{-2t/n} (\prod_i X_i)^2$
largely cancels out the $(\prod_i X_i)^2$ in the denominator, leaving an
expression which (setting aside questions of counting and degeneracy of
vacua, as well as how the change of variables would operate on
kinetic terms) formally duplicates the usual expression in the orbifold:
\begin{displaymath}
\langle {\cal O}_1 \cdots {\cal O}_n \rangle_{\rm quintic} \: \propto \:
\sum_{\rm vacua} \frac{ {\cal O}_1 \cdots {\cal O}_n }{
 \det\left(\partial_{X_i} 
\partial_{X_j} W \right) 
}.
\end{displaymath}

Now, setting aside formal justifications, it is, of course,
tempting to conjecture a (0,2) analogue of the
story above.  The justification for the (2,2) case in 
\cite{Aganagic:2004yn} was based on the claim in 
\cite{Schwarz:1995ak} that the A model on a hypersurface is equivalent
to the A model on a supermanifold over the ambient space.  Unfortunately,
at present no (0,2) analogue of \cite{Schwarz:1995ak} exists in the
literature.  Mathematically, the idea of inserting Chern classes to
restrict to hypersurfaces is more properly understood in terms of
insertions of Mathai-Quillen forms.  A (0,2) analogue of Mathai-Quillen
forms has been sketched out (see \cite{Garavuso:2013zoa} 
for definitions and conjectures), but one 
open question in that work is whether they have all the pertinent properties
of ordinary Mathai-Quillen forms.

Another question concerns the appropriateness of getting a mirror at a 
Landau-Ginzburg orbifold point.  In (2,2) theories, for hypersurfaces,
since the B model is independent of K\"ahler structures, we could 
evaluate a B-twisted mirror at any convenient point on the K\"ahler
moduli space, and the Landau-Ginzburg orbifold point is a convenient
such point.  By contrast, (0,2) theories are somewhat more complicated.
It was argued in 
\cite{McOrist:2008ji,Donagi:2011uz,Donagi:2011va,Closset:2015ohf} 
that A/2-twisted GLSMs are independent of
complex structure moduli and some bundle moduli ($J$s), and B/2-twisted
GLSMs are independent of K\"ahler moduli and other bundle moduli ($E$s),
but this seems to be an accident of the GLSM's presentation of moduli,
and we do not understand how to formulate analogous statements in
IR nonlinear sigma models -- or indeed if it is even possible to formulate
such statements in IR theories.
In any event, for our purposes, this will suffice to justify describing
B/2-twisted mirrors at Landau-Ginzburg orbifold points.

With all that in mind, we will now describe some formal computations for
(0,2) theories,
generated by proceeding along the same lines as (2,2) theories, albeit
with less justification.

Note that in (0,2) theories, in addition to performing a variable change
on (0,2) chiral superfields, we must also perform a variable change on
(0,2) Fermi superfields.  For example, suppose that in the mirror
to the ambient GLSM, the fundamental fields are $(\theta_A, G_A)$,
where $\theta_A$s are (0,2) chiral superfields and $G_A$s are (0,2)
Fermi superfields, which on the (2,2) locus together form a (2,2) chiral
multiplet.  If we define
\begin{displaymath}
X_A \: = \: \exp(-\theta_A/n),
\end{displaymath}
and take that to be a fundamental chiral superfield
then to get a Landau-Ginzburg model that matches the (2,2)
mirror on the (2,2) locus, we must 
also define new fundamental Fermi superfields $\Lambda_A$, as
\begin{displaymath}
\Lambda_A \: = \: \frac{\partial X_A}{\partial \theta_B} G_B.
\end{displaymath}
In the example above,
\begin{displaymath}
\frac{\partial X_A}{\partial \theta_B} \: = \: 
- \frac{1}{n} X_A \delta^B_A,
\end{displaymath}
hence
\begin{displaymath}
\Lambda_A \: = \: - \frac{1}{n} X_A G_A
\mbox{   (no sum on $A$)},
\end{displaymath}
and we should take $\Lambda_A$ to be the fundamental field in the
mirror to the hypersurface, replacing
$G_A$.

\subsection{Example}

For an example, we will study the mirror to a (0,2) GLSM describing
a degree $(n,0)$ hypersurface in ${\mathbb P}^2 \times {\mathbb P}^1$,
with a tangent bundle deformation.
There are two dueling constraints that make finding interesting examples,
somewhat nontrivial:
\begin{itemize}
\item Our $E$'s, $J$'s must satisfy $E \cdot J = 0$ in order to preserve
supersymmetry,
\item Simultaneously, choices of $S \subset Q$ constrain the allowed
deformations.
\end{itemize}

We will deal with these issues by considering a hypersurface in the
first ${\mathbb P}^2$ factor, paired with a bundle deformation over the
second ${\mathbb P}^1$ factor, so that the two constraints are
fundamentally uncoupled from one another.
We will `follow our nose' and work out what naive expectations would
predict for the (0,2) mirror given the proposal of this paper; 
however, we have not checked any
correlation functions or performed
other independent tests to ensure that we have
in fact produced the correct (0,2) mirror.

Our GLSM will have six chiral superfields, which we label $\phi_{1,\cdots,5}$
and $p$, with charge matrix
\begin{displaymath}
Q \: = \: \left[ \begin{array}{cccccc}
1 & 1 & 1 & 0 & 0 & -n \\
0 & 0 & 0 & 1 & 1 & 0
\end{array} \right].
\end{displaymath}
We interpret $\phi_{1-3}$ as related to homogeneous coordinates on 
first factor (${\mathbb P}^2$), and $\phi_{4-5}$ as related to homogeneous
coordinates on the second factor (${\mathbb P}^1$).

Given the charge matrix $Q$ above, we will take
\begin{displaymath}
(V^A_i) \: = \: \left[ \begin{array}{cccccc}
n & 0 & 0 & 0 & 0 & 1 \\
0 & 1 & 0 & 0 & 0 & 1/n \\
0 & 0 & 1 & 0 & 0 & 1/n \\
0 & 0 & 0 & -1 & 1 & 0
\end{array} \right].
\end{displaymath}

Now, since we are going to apply our mirror construction, we pick an
invertible submatrix $S \subset Q$.  To be specific, we will take $S$ to
correspond to the fourth and sixth rows of $Q$, so that 
\begin{displaymath}
S \: = \: \left[ \begin{array}{cc}
0 & -n \\
1 & 0 \end{array} \right],
\end{displaymath}
and we will consider deformations
\begin{displaymath}
(A_{ij}) \: = \: \left[ \begin{array}{cccccc}
1 & 0 & 0 & 0 & 0 & 0 \\
0 & 1 & 0 & 0 & 0 & 0 \\
0 & 0 & 1 & 0 & 0 & 0 \\
0 & 0 & 0 & 1 & 0 & 0 \\
A_{51} & A_{52} & A_{53} & A_{54} & A_{55} & A_{56} \\
0 & 0 & 0 & 0 & 0 & 1
\end{array} \right].
\end{displaymath}
Now, the choice of $S$ above is compatible with a more complicated 
deformation matrix $(A_{ij})$ -- 
for example, the first, second, and third rows could have entries
different from those of the identity matrix.  However, to avoid running
into difficulties with the constraint $E \cdot J = 0$, for simplicity
in this example we pick the $E$'s to be trivial along directions with
nonzero $J$'s, and so we take the matrix to be of the more specialized
form above.

With these choices, it is straightforward to compute
\begin{displaymath}
(D^A_{i_S}) \: = \:  \left[ \begin{array}{cc} 
0 & 0 \\
0 & 0 \\
0 & 0 \\
(A_{54}+A_{55}-1)/n &
- ( A_{51}+A_{52}+A_{53} - n A_{56})
\end{array}
\right].
\end{displaymath}

Thus, the (0,2) mirror to the GLSM {\it without superpotential} is given by
a Landau-Ginzburg model with superpotential
\begin{eqnarray*}
W & = & - G_1 \big( n \exp(-Y_1) + \exp(-Y_p) \big) \: - \:
G_2 \big( \exp(-Y_2) + (1/n) \exp(-Y_p) \big) \\
& &   - 
G_3 \big( \exp(-Y_3) + (1/n) \exp(-Y_p) \big) \\
& & -G_4 \bigg( - \exp(-Y_4) + \exp(-Y_5) + \frac{1}{n}(A_{54}+A_{55}-1)
\exp(-Y_4)
\\
& & \hspace*{0.6in}
 - (A_{51}+A_{52}+A_{53}-nA_{56}) \exp(-Y_p)
\bigg),
\end{eqnarray*}
subject to the usual D-term constraints, which allow us to rewrite
$Y$'s in terms of $\theta$'s.  If we pick 
\begin{displaymath}
(\tilde{t}_i) \: = \: (0,0,0,t_2,0,- t_1/n ),
\end{displaymath}
then
\begin{displaymath}
\theta_1 = Y_1/n, \: \: \:
\theta_2 = Y_2, \: \: \:
\theta_3 = Y_3, \: \: \:
\theta_4 = Y_5,
\end{displaymath}
and
\begin{displaymath}
Y_p =  \theta_1 + \theta_2/n +  \theta_3/n - t_1/n, \: \: \:
Y_4 = - \theta_4 + t_2,
\end{displaymath}
so if we define 
\begin{displaymath}
Z_1 = \exp(-\theta_1), \: \: \:
Z_{2,3} = \exp(-\theta_{2,3}/n), \: \: \:
Z_4 = \exp(-\theta_4),
\end{displaymath}
the mirror superpotential above becomes
\begin{eqnarray*}
W & = & - G_1 \left( n Z_1^n + Z_1 Z_2 Z_3 e^{-t_1/n} \right) \: - \:
G_2 \left( Z_2^n + \frac{1}{n} Z_1 Z_2 Z_3 e^{-t_1/n} \right) \\
& & - G_3 \left(  Z_3^n + \frac{1}{n} Z_1 Z_2 Z_3 e^{-t_1/n} \right)
 \\
& & - G_4 \left( Z_4 - \frac{q_2}{Z_4} + \frac{1}{n}\left( A_{54}+A_{55}-1
\right) \frac{q_2}{Z_4} - 
\left( A_{51}+A_{52}+A_{53}-nA_{56} \right) Z_1 Z_2 Z_3 e^{-t_1/n}
 \right),
\end{eqnarray*}
and the theory has a $({\mathbb Z}_n)^2$ orbifold, acting on $Z_{2,3}$,
in which the group
action preserves the superpotential above.

For completeness, the operator mirror map takes the following form:
\begin{eqnarray*}
\exp(-Y_1) = Z_1^n & \leftrightarrow & \sigma_1, \\
\exp(-Y_2) = Z_2^n & \leftrightarrow & \sigma_1,\\
\exp(-Y_3)  = Z_3^n & \leftrightarrow & \sigma_1, \\
\exp(-Y_4) = \frac{q_2}{Z_4} & \leftrightarrow & \sigma_2, \\
\exp(-Y_5) = Z_4 & \leftrightarrow & (A_{51}+A_{52}+A_{53}-nA_{56}) \sigma_1 +
(A_{54}+A_{55}) \sigma_2, \\
\exp(-Y_p) = Z_1 Z_2 Z_3 e^{-t_1/n}  & \leftrightarrow & -n \sigma_1.
\end{eqnarray*}

We could also add a twisted mass along the $p$ direction, for example.
Doing so would add the following terms to the
mirror Landau-Ginzburg model superpotential
\begin{displaymath}
G_A V_p^A \tilde{m}_p \: = \: \tilde{m}_p \left( G_1 + \frac{1}{n} G_2
+ \frac{1}{n} G_3 \right),
\end{displaymath}
and also
alter the operator mirror map, by modifying the map for $\exp(-Y_p)$ as
\begin{displaymath}
\exp(-Y_p) \: \leftrightarrow \: -n \sigma_1 + \tilde{m}_p.
\end{displaymath}

Now, let us consider restricting to the hypersurface.  This will involve
changing the fundamental fields, from $\theta$'s to $X$'s 
and $G$'s to $\Lambda$'s.

First, let us consider changing variables amongst chiral superfields.
We will take $Z_i$s to be the fundamental variables.
(Physically, this means changing {\it e.g.} kinetic terms, and so
changing the physical theory, but here we will primarily focus on the
superpotential.)

In addition, we must also change the Fermi fields, as mentioned
previously.  Following the pattern discussed previously, we define
the new fundamental Fermi superfields
\begin{displaymath}
\Lambda_1 = - Z_1 G_1, \: \: \:
\Lambda_{2,3} = - \frac{1}{n} Z_{2,3} G_{2,3}, \: \: \:
\Lambda_4 = - Z_4 G_4.
\end{displaymath}

Rewriting the mirror superpotential above in terms of the new
fundamental fields $Z_i$, $\Lambda_i$, we find it takes the form
\begin{eqnarray*}
W & = & \Lambda_1\left( n Z_1^{n-1} + Z_2 Z_3 e^{-t_1/n} \right)
\: + \: \Lambda_2 \left( n Z_2^{n-1} + Z_1 Z_3 e^{-t_1/n} \right) \\
& & + \Lambda_3 \left( n Z_3^{n-1} + Z_1 Z_2 e^{-t_1/n} \right) \\
& & + \Lambda_4\left( 1 - \frac{q_2}{Z_4^2} + \frac{1}{n}
\left( A_{54}+A_{55}-1 \right) \frac{q_2}{Z_4^2} - 
\left( A_{51}+A_{52}+A_{53}-nA_{56} \right) \frac{Z_1 Z_2 Z_3}{Z_4}
e^{-t_1/n} \right),
\end{eqnarray*}
again with a $({\mathbb Z}_n)^2$ orbifold group action.

The reader should note that the first three terms in the superpotential
above appear identical to those one would expect in a (0,2) expansion
of a (2,2) superpotential of the form
\begin{displaymath}
Z_1^n + Z_2^n + Z_3^n + Z_1 Z_2 Z_3 e^{-t_1/n},
\end{displaymath}
exactly as one would expect for the mirror to a hypersurface of
degree $n$ in ${\mathbb P}^2$.
Similarly, the first two terms in the last line appear to be a (0,2)
expansion of the (2,2) Toda dual to ${\mathbb P}^1$, defined by the
superpotential
\begin{displaymath}
Z_4 + \frac{q_2}{Z_4},
\end{displaymath}
again exactly as one would expect.  The remaining terms encode the
bundle deformation.

\subsection{Brief comment on the quintic}

Next, let us consider an example of a different character.
Consider a GLSM for a quintic in ${\mathbb P}^4$.
As is well-known (see {\it e.g.} \cite{Green:1987mn}[section 15.6.3]), 
one can deform it to a (0,2) theory.
To see this, first rewrite the (2,2)superpotential in (0,2) language as 
\begin{displaymath}
W \: = \: \sum_i \Lambda^i J_i \: + \: \Lambda^p J_p,
\end{displaymath}
for
\begin{displaymath}
J_i \: = \: p \frac{ \partial G}{\partial \phi_i }, \: \: \:
J_p \: = \: G,
\end{displaymath}
for $G$ a quintic polynomial in the chiral superfields $\phi_i$.
Then, one typically deforms off the (2,2) locus by deforming
\begin{displaymath}
J_i \: = \: p \frac{ \partial G}{\partial \phi_i }
\: \mapsto \: p \frac{ \partial G}{\partial \phi_i } \: + \:
G_i,
\end{displaymath}
where the functions $G_i$ are constrained to obey
\begin{displaymath}
\sum_i \phi_i G_i \: = \: 0,
\end{displaymath}

Now, let us consider this theory in the context of the mirror proposals
of this paper.  The A/2-twist only depends upon $E$'s, not $J$'s, so the
deformation above is invisible in the twisted theory.  The dualization
procedure we have described does not seem to involve the $G_i$'s above --
we take a tangent bundle deformation on the ambient space and restrict
to a hypersurface, but the $G_i$'s cannot be understood as a tangent
bundle deformation of the ambient space.  As a result, using the methods
here, the $G_i$'s are effectively invisible.  On the other hand, since
the A/2 theory is independent of the $G_i$'s, it is nevertheless
a sensible mirror for the A/2 theory.

\section{Conclusions}

In this paper we have given a proposal for (0,2) mirrors to toric
Fano varieties with special tangent bundle deformations, subsets of
toric deformations, and also described
restrictions to hypersurfaces.  Our methods do not apply to all
tangent bundle deformations, only a subset of the toric deformations.
We have given formal arguments that the resulting correlation functions
always match, and also checked in examples that our methods reproduce
previous results for (0,2) mirrors produced by laboriously guessing
and tuning ansatzes.

It would be interesting to generalize the results presented here to
all tangent bundle deformations.  On the one hand, for more general
tangent bundle deformations, previous methods have generated nonlinear
superpotential terms in the mirror, which we do not see.  On the other
hand, naive moduli counting arguments suggest that in many cases it
may be possible to use field redefinitions to rewrite all the 
GLSM-realizable tangent bundle deformations in the form we use in this paper.

\section{Acknowledgements}

We would like to thank L.~Anderson, C.~Closset, J.~Gray,
I.~Melnikov, and R.~Plesser for useful
conversations.  E.S. was partially supported by NSF grant PHY-1417410.

\appendix

\section{Brief notes on (2,2) mirror ansatz}
\label{app:alt}

In this appendix we will
briefly outline how symmetries and the operator mirror map partially
determine the exponential terms in the (2,2) GLSM mirror superpotential.
Suppose we have not derived the
instanton-generated terms, and only have an ansatz for
the mirror superpotential of the form
\begin{equation}  \label{eq:mirror-ansatz}
W \: = \: \sum_{a=1}^k \Sigma_a \left( \sum_{i=1}^NQ^a_i Y_i - t_a\right) \: + \:
g(Y_i),
\end{equation}
for some unknown function $g(Y_i)$.  (Requiring R-charges match only fixes
terms $\exp(-Y_i)$ up to an R-invariant function.)
Instead of deriving $g$ from a direct
instanton computation in the A-twisted theory, we outline here how the 
same result could be obtained using other properties of the theory.

Now, previously we derived
the operator mirror 
map~(\ref{eq:22mirror-basic}) from the form of the mirror superpotential,
but 
one can outline an independent justification,
and then use it
to demonstrate the form of $g$.  To see this, use the relation
\cite{Hori:2000kt}[equ'n (3.17)]
\begin{displaymath}
Y + \overline{Y} \: = \: 2 \overline{\Phi} e^{2QV} \Phi,
\end{displaymath}
which implies component relations \cite{Hori:2000kt}[equ'ns (3.20), (3.21)]
\begin{displaymath}
\chi_+ = 2 \overline{\psi}_+ \phi, \: \: \:
\overline{\chi}_- = - 2 \phi^{\dag} \psi_-,
\end{displaymath}
for $\chi$ the superpartners of $Y$ and $\psi$ the superpartners of
$\phi$.  From \cite{Witten:1993yc}[equ'n (2.19)], the equations of
motion of $\overline{\sigma}_a$ (in the limit $e^2 \rightarrow \infty$,
so that the kinetic terms drop out) are
\begin{displaymath}
\left( \sum_{i=1}^N Q_i^a | \phi_i |^2 \right) \sigma_a \: \propto \:
\sum_{i=1}^N  \overline{\psi}_{+ i} \psi_{- i}.
\end{displaymath}
If we add twisted masses so that only one $\phi$ field
is light, then this becomes
\begin{displaymath}
Q_i^a \sigma_a + \tilde{m}_i \: \propto \: \frac{\overline{\psi}_{+ i} \psi_{- i}}{
| \phi_i|^2 } \: \propto \:
\chi_{+ i} \overline{\chi}_{- i}.
\end{displaymath}
Now, the $\chi_+ \overline{\chi}_-$ could come from a $Y^2$, but that
has the wrong R charge to make the expression sensible.  However,
$\exp(-Y_i)$ has the correct R charge and contains $Y^2$, so up to
overall factors, which can be reabsorbed into field redefinitions,
this suggests
\begin{displaymath}
Q_i^a \sigma_a + \tilde{m}_i \: = \: \exp(-Y_i),
\end{displaymath}
which is the operator mirror map~(\ref{eq:22mirror-basic}).
(Granted, we are again using axial R-charges, but here since we know
some components, there is less ambiguity.)

Returning to the ansatz~(\ref{eq:mirror-ansatz}), 
we can now determine the function $g$.
The equations of motion from the superpotential above imply
\begin{eqnarray*}
\frac{\partial W}{\partial Y_i} & = &  Q_i^a \sigma_a \: + \:
\frac{\partial g}{\partial Y_i}, \\
& = & 0,
\end{eqnarray*}
and the operator mirror map implies
\begin{displaymath}
Q_i^a \sigma_a + \tilde{m}_a \: = \: \exp(-Y_i),
\end{displaymath}
hence
\begin{displaymath}
\frac{\partial g}{\partial Y_i}
\: = \: - Q_i^a \sigma_a \: = \: \tilde{m}_i  - \exp(-Y_i),
\end{displaymath}
hence
\begin{displaymath}
g(Y_i) \: = \: \tilde{m}_i Y_i + \sum_i \exp(-Y_i),
\end{displaymath}
up to an irrelevant additive constant.

It is tempting to apply the same methods to (0,2) theories. Unfortunately,
the decreased symmetry leads to multiple possible potential (0,2)
mirror superpotentials, derived from applying the operator mirror map
in different ways,
which must be independently
tested against chiral rings and correlation functions.

\end{document}